%% file: paper.tex
\documentclass[lettersize,journal]{IEEEtran}

\usepackage[english]{babel}
\usepackage{amsmath,amsfonts} 
\usepackage{amssymb}
\usepackage{amsthm}
\usepackage{graphicx}
\usepackage{multirow}
\usepackage{cite}
\usepackage{url} 
\usepackage[colorlinks=true,linkcolor=black,citecolor=black,pdfborder={0 0 0}]{hyperref}
\usepackage{color}
\usepackage{enumitem}
\usepackage[kerning,spacing]{microtype}
\usepackage{algorithm}
\usepackage{array} 
\usepackage{stfloats} 
\usepackage{verbatim}
\usepackage{algpseudocode}
\usepackage{multicol}
\usepackage{threeparttable}
\usepackage{xspace}
\usepackage{pifont}
\usepackage[font=small,labelfont=bf]{caption}
\usepackage{booktabs}
\usepackage{textcomp}
\usepackage{xcolor}
\usepackage{subcaption}
\usepackage{balance}
\usepackage{todonotes}
\usepackage{makecell}

\usepackage{mathptmx}
\DeclareMathAlphabet{\mathcal}{OMS}{cmsy}{m}{n}
\DeclareMathAlphabet{\mathrm}{OT1}{bch}{m}{n}
\DeclareMathAlphabet{\mathit}{OT1}{bch}{m}{it}

\algnewcommand{\LineComment}[1]{\State $\triangleright$ #1}

\setitemize{noitemsep,topsep=3pt,parsep=0em}

\DisableLigatures[f]{encoding=*, family=*}

\newcommand{\cmark}{\ding{51}}
\newcommand{\xmark}{\ding{55}}
\newcommand{\sysname}{Fletch\xspace}
\newcommand{\sysnameplus}{Fletch+\xspace}

\definecolor{OliveGreen}{cmyk}{0.64,0,0.95,0.40}
\definecolor{ao}{rgb}{0.0, 0.5, 0.0}
\definecolor{asparagus}{rgb}{0.53, 0.66, 0.42}
\definecolor{applegreen}{rgb}{0.55, 0.71, 0.0}
\definecolor{aogreen}{rgb}{0.0, 0.5, 0.0}
\definecolor{columbiablue}{rgb}{0.61, 0.87, 1.0}
\definecolor{cornellred}{rgb}{0.7, 0.11, 0.11}
\definecolor{cornflowerblue}{rgb}{0.39, 0.58, 0.93}
\definecolor{denim}{rgb}{0.08, 0.38, 0.74}

\colorlet{BtfulGreen}{black!30!green!70!}
\colorlet{BtfulOrange}{white!10!orange!90!}
\colorlet{BtfulGray}{white!50!gray!50!}

\newcommand{\para}[1]{\noindent\textbf{#1}}
\newcommand{\cut}[1]{}

\newcommand*\circled[1]{\tikz[baseline=(char.base)]{
\node[shape=circle,draw,inner sep=0.6pt] (char) {\small {#1}};}}

\begin{document}

\title{\sysname: File-System Metadata Caching in Programmable Switches}

\author{Qingxiu Liu, Jiazhen Cai, Siyuan Sheng, Yuhui Chen, Lu Tang, Zhirong Shen, and Patrick P. C. Lee
\thanks{This work was supported in part by National Natural Science Foundation of China (62302410, U25B2037) and Research Grants Council of Hong Kong (GRF
14201523). Corresponding author: Patrick P. C. Lee.}
\thanks{Qingxiu Liu, Jiazhen Cai, Siyuan Sheng, and Patrick P. C. Lee are with
the Department of Computer Science and Engineering, The Chinese University of
Hong Kong, Hong Kong (e-mail: qxliu23@cse.cuhk.edu.hk; jzcai@cse.cuhk.edu.hk;
siyuanshengpro@gmail.com; pclee@cse.cuhk.edu.hk).}
\thanks{Yuhui Chen, Lu Tang and Zhirong Shen are with the Department of
Computer Science and Technology, Xiamen University, Xiamen 361005, China
(e-mail: yhchen230@stu.xmu.edu.cn; tanglu@xmu.edu.cn;
zhirong.shen2601@gmail.com).}}


\maketitle

\begin{abstract}
Fast and scalable metadata management across multiple metadata servers is crucial for distributed file systems to handle numerous files and directories. Client-side caching of frequently accessed metadata can mitigate server loads, but it incurs significant overhead and complexity in maintaining cache consistency when the number of clients increases.  
We propose \sysname, an in-switch file-system metadata caching framework that leverages programmable switches to serve file-system metadata requests from multiple clients directly in the switch data plane. Unlike prior in-switch key-value caching systems, \sysname addresses file-system-specific path dependencies under stringent switch resource constraints.  We implement \sysname atop HDFS and evaluate it on a Tofino-switch testbed using real-world file-system metadata workloads.  \sysname achieves up to 181.6\% higher throughput than vanilla HDFS and complements client-side caching with throughput gains of up to 139.6\% on 128 simulated metadata servers.  
\end{abstract}


\section{Introduction}
\label{sec:introduction}

Scaling file-system metadata management across multiple metadata servers is crucial for distributed file systems to handle billions of files and directories. Field studies, both classical \cite{roselli00,harter12} and recent \cite{wang23}, indicate that metadata operations dominate file-system requests. For example, 67-96\% of file-system requests in Baidu AI Cloud are
metadata-related \cite{wang23}.  As the number of files and directories grows, especially in workloads dominated by small files \cite{carns09,harter14}, metadata operations, such as inode lookups, permission checks, and directory traversals, become performance bottlenecks. File-system metadata access patterns in practice tend to be highly skewed, with a small fraction of files and directories accessed far more frequently than others \cite{leung08,abad12,ananthanarayanan11}; for example, less than 3\% of files account for 34-39\% of requests in Yahoo's HDFS clusters \cite{abad12}. Metadata servers are frequently bottlenecked by skewed access, leading to severe load imbalance where the entire cluster's performance is limited by a few overloaded servers.

Caching frequently accessed metadata on the client side is a common strategy to mitigate loads on metadata servers. While client-side caching distributes the caching load across clients, it introduces a severe distributed cache consistency problem: maintaining consistency across numerous clients requires complex, high-overhead lease management or invalidation protocols \cite{welch08,ren14,li17,lv22}, since any metadata update must notify all relevant clients. Maintaining this consistent state requires metadata servers to track every participant's cache and broadcast invalidations upon updates, resulting in a communication complexity of $O(N)$, where $N$ is the number of clients. This overhead incurs high latency and makes client-side caching increasingly fragile in dynamic environments with frequent client churn. More fundamentally, because client-side caches are distributed across individual clients, they cannot coordinate to resolve the severe load imbalance at the backend servers caused by highly skewed file-system access patterns. Existing client-side caching approaches \cite{welch08,ren14,li17,lv22} further limit their scope to caching directory permission metadata to mitigate path resolution overhead (\S\ref{subsec:metadata}) while still relying on metadata servers for attribute retrieval, so their overall performance gains remain modest.

Programmable switches  \cite{bosshart13} offer a compelling and unique alternative by leveraging their fundamentally centralized network position to enable {\em in-switch caching}. Since the switch sits on the critical path between all clients and servers, it maintains a global view of all metadata requests, a structural vantage point that effectively addresses the limitations of client-side caching. First, since cache updates are applied directly within the network rather than coordinated across distributed clients, in-switch caching natively eliminates the need for inter-client invalidation protocols, reducing the complexity of cache consistency maintenance from $O(N)$ to $O(1)$. Second, in-switch caching absorbs highly skewed, hot-path requests directly in the switch data plane, thereby effectively shielding backend servers from load imbalance.

While prior work has explored in-switch caching for key-value stores \cite{li16,jin17,liu19,sheng25,li20,eldakiky20,lee21}, in-switch caching for file-system metadata poses unique challenges that have not been directly addressed before. First, file-system pathnames, unlike fixed-length keys, have large, variable sizes and cannot readily fit within limited switch resources. Second, accessing a file's metadata requires traversing the metadata of its internal directories, thereby exacerbating switch resource demands when caching multiple levels of metadata. Third, cache lookups for file and directory metadata require multiple iterations under the strict switch programming model, complicating concurrent cache updates and lookups while maintaining consistency and performance.

We propose \sysname, an in-switch \underline{f}i\underline{l}e-system m\underline{et}adata ca\underline{ch}ing framework tailored for distributed file systems under skewed, read-intensive workloads. \sysname extends in-switch key-value caches by specifically addressing file-system path dependencies through various techniques: (i) \emph{path-aware cache management} to account for path dependencies in cache admission and eviction; (ii) \emph{multi-level read-write locking} to enable high-performance concurrent cache lookups and updates; and (iii) \emph{local hash collision resolution} to efficiently and correctly map variable-length paths into fixed-length keys. 
It targets highly skewed, read-intensive file-system workloads and supports two deployment modes (\S\ref{subsec:modes}), a {\em standalone} mode (replacing client-side caching, eliminating $O(N)$ consistency overhead) and a {\em complementary} mode (augmenting client-side caching with in-switch load balancing), for distinct deployment regimes.

We implemented \sysname in P4 \cite{bosshart14} and compiled it into the Tofino switch chipset \cite{tofino,switchthpt}. We integrated \sysname with Hadoop HDFS \cite{hdfs}, while preserving HDFS semantics.  We also implemented a state-of-the-art client-side caching approach \cite{ren14,lv22} to demonstrate how \sysname complements it and further improves file-system metadata access performance.  Our evaluation on four real-world workloads \cite{ren14,lv22,xu24} shows that with 128 simulated metadata servers, \sysname achieves up to 181.6\% higher throughput than vanilla HDFS, while client-side caching coupled with \sysname achieves up to 139.6\% higher throughput than without \sysname.
We release the source code of \sysname at \url{https://github.com/adslabcuhk/fletch}.

\section{Background and Motivation}
\label{sec:background}

\subsection{File-System Metadata Management}
\label{subsec:metadata}

File systems organize data in a hierarchical, tree-based namespace, where
files and directories are managed as leaf and non-leaf nodes, respectively.
Each node is identified by a {\em path} and contains {\em metadata} (e.g.,
owner, size, and permission).  A path often has multiple internal directories.
We refer to a file or directory below the root as the $i$-th {\em level}
($i\ge 1$), and the maximum number of levels of a path as the {\em depth}.
For example, the path \texttt{/a/b/c.txt} has a depth of 3, with levels 
\texttt{/a}, \texttt{/a/b}, and \texttt{/a/b/c.txt}.  In the namespace, we say
that path $p$ is an {\em ancestor} of path $q$, and $q$ is a {\em descendant}
of $p$, if $p$ is an internal directory (or prefix) of $q$.  We say that $p$
is the {\em parent} of $q$, and $q$ is a {\em child} of $p$, if $p$ is an
ancestor of $q$ with exactly one less level than $q$ (e.g., \texttt{/a} is the
parent of \texttt{/a/b}, and \texttt{/a/b} is a child of \texttt{/a}).

File-system metadata management is crucial for distributed file systems. There are various metadata operations for files (e.g., \texttt{create}, \texttt{delete}, \texttt{open}, and \texttt{close}), directories (e.g., \texttt{mkdir} and \texttt{rmdir}), attributes (e.g., \texttt{chmod}, \texttt{chown}, and \texttt{utime}), and data organization (i.e., \texttt{rename}, \texttt{readdir}, and \texttt{stat}). File-system metadata operations often rely on {\em path resolution}, which parses and traverses each level of a path to verify metadata for existence and permissions.

\begin{figure}[!t]
\centering
\includegraphics[width=0.85\linewidth]{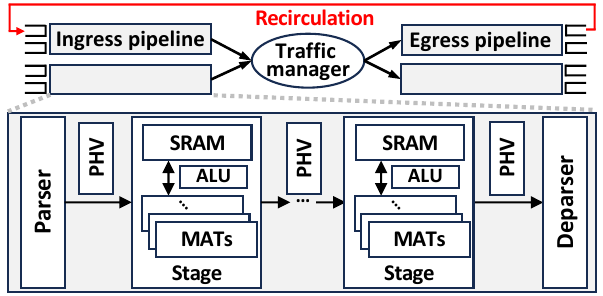}
\caption{Data plane of a programmable switch.}
\label{fig:switch}
\vspace{-12pt}
\end{figure}

\subsection{Programmable Switches}
\label{subsec:switch}

Programmable switches \cite{bosshart13} enable customized packet processing
tasks beyond traditional packet forwarding. 
A programmable switch comprises a data plane and a control plane.  The data
plane, as shown in Figure~\ref{fig:switch}, performs packet forwarding via
multiple {\em ingress} and {\em egress} pipelines.  Packets enter an ingress
pipeline and are forwarded to an egress pipeline via a {\em traffic manager}.
The control plane manages the data plane by specifying packet processing
rules. 

Each ingress or egress pipeline processes packets via a series of {\em
stages}, each of which contains {\em match-action tables (MATs)} that execute
processing logic based on the Reconfigurable Match-Action Table (RMT) paradigm \cite{bosshart13}.
A {\em parser} converts packet headers into {\em packet
header vectors (PHVs)}, processed by MATs with on-chip ALUs across
stages.  Each stage can only access limited SRAM, typically with tens of
memory blocks.  The switch programming model imposes strict constraints: each
memory block can be accessed at most once per PHV traversal, and stages cannot
access memory blocks of other stages.  After processing, a {\em deparser}
reconstructs packets.  To process packets in multiple
iterations, switches support {\em recirculation}, which redirects packets from
an egress pipeline back to an ingress pipeline. Recirculation should be
cautiously used, as it consumes extra switch resources and slows down packet
forwarding. In this work, we leverage recirculation for read requests that
require in-switch path resolution (\S\ref{subsec:pathhashing}) and write
requests that await locks (\S\ref{subsec:lockflow}).

\subsection{Challenges}
\label{subsec:challenges}

Designing in-switch metadata caching for distributed file systems poses
the following challenges. 

\para{Challenge 1: Constrained switch resources.}  Switches have
limited resources and complicate caching implementation.  For example, Tofino
switches \cite{tofino} provide only 12~stages per ingress or egress pipeline, 
each processing up to 16~bytes due to ALU word size restrictions.  The PHV
size is capped (e.g., 768~bytes in Tofino switches \cite{tofino}), and all 
pipelines share limited SRAM.  For comparison, NetCache \cite{jin17}
allocates SRAM for access frequency tracking and supports only a maximum value
size of 128~bytes for key-value records.  

File-system metadata caching is particularly resource-intensive. HDFS file or
directory names can reach 255~bytes \cite{defaultsetting} and aggravate SRAM
and PHV overhead.  Path resolution for a path (e.g., \texttt{/a/b.txt})
requires accessing metadata for its ancestors (i.e., \texttt{/} and
\texttt{/a}) (\S\ref{subsec:metadata}).  Caching metadata for all paths and
their ancestors is desirable but challenging given limited SRAM. Splitting
path processing across multiple levels requires careful synchronization across 
stages.  To mitigate switch resource usage, existing in-switch key-value
caches \cite{jin17,liu19,sheng25} offload cache admission and eviction to a
centralized controller, but excessive switch-to-controller communications
incur high latencies. 

\para{Challenge 2: Cache consistency.}  Enforcing cache consistency under
concurrent updates is crucial, as programmable switches handle requests from
multiple ingress pipelines and trigger simultaneous cache updates.  Cache
updates may overlap with path resolution, thereby further complicating cache
consistency management.  For example, a read request for \texttt{/a/b.txt}
requires metadata access for \texttt{/}, \texttt{/a}, and \texttt{/a/b.txt},
while concurrent \texttt{chmod} requests for \texttt{/a} and \texttt{/a/b.txt}
update their metadata and may occur between their cache lookups.  Thus, the
read request can access a mix of pre- and post-updated metadata, leading to
inconsistencies. 

\section{Design Overview}
\label{sec:overview}

\subsection{Goals and Assumptions}
\label{subsec:goals}

\sysname is an in-switch metadata caching framework for distributed file
systems, aiming to achieve high throughput and low latencies for file-system
metadata requests across multiple metadata servers.
It targets read-intensive file-system metadata workloads, as observed in
prior studies \cite{ren14,lv22,xu24}.  For example, in a LinkedIn HDFS
cluster, 84\% of 145~million metadata operations are lookups, with only 9\%
creates and 7\% updates \cite{ren14}; in Alibaba's Pangu file system, over
60\% of metadata operations are reads \cite{lv22}.

To ensure consistency, \sysname adopts {\em write-through} caching, which synchronously updates both the in-switch cache and server-side metadata before acknowledging clients, as in prior in-switch key-value caches \cite{li16,liu17,jin17,liu19}, rather than write-back caching. While write-back caching reduces write latency by buffering updates in the switch data plane and immediately acknowledging clients, this benefit is marginal for read-intensive file-system metadata workloads, which \sysname targets, since writes constitute only a small fraction of all requests, and the aggregate impact of high per-write latency on overall throughput is limited. Most importantly, under write-back caching, the switch data plane holds the sole up-to-date copy of new metadata until it is written back to the server; if the switch fails before this write-back completes, the new metadata is lost. Write-through caching eliminates this risk by ensuring that metadata servers always hold the latest committed state, providing durability independent of switch failures.

Since path resolution always starts at the root, \sysname further assumes that the root directory's metadata is persistently cached, with its permissions unchanged and the root never deleted \cite{niazi17}, consistent with standard distributed file-system practices. 

\subsection{Architecture and Design Roadmap}
\label{subsec:architecture}

\begin{figure}[!t]
\centering
\includegraphics[width=0.99\linewidth]{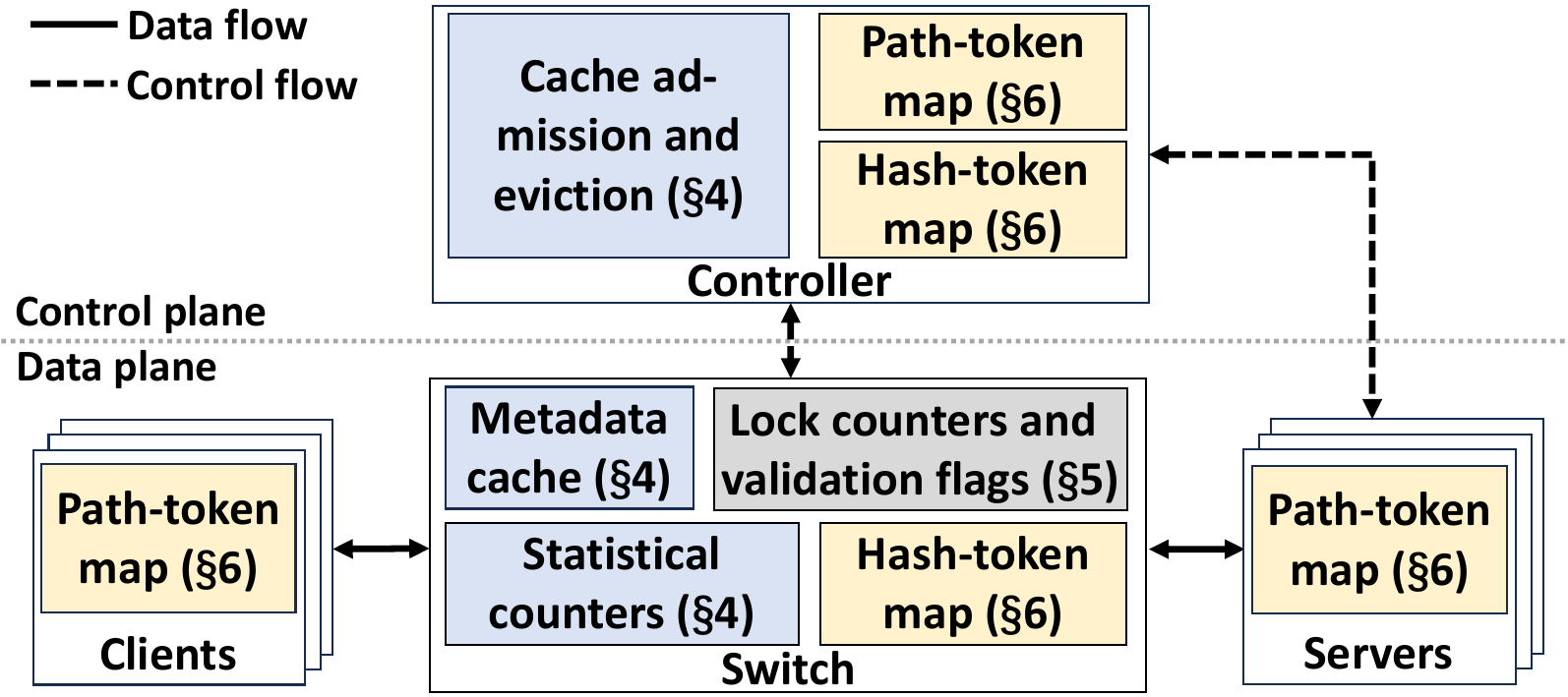}
\caption{\sysname's architecture.}
\label{fig:architecture}
\vspace{-9pt}
\end{figure}

Figure~\ref{fig:architecture} depicts \sysname's architecture. 
In the control plane, a
controller manages in-switch cache admission and eviction, while in the data
plane, multiple clients send (receive) file-system metadata requests
(responses) via the switch to (from) multiple metadata servers (or servers for
short).  \sysname employs three techniques to enable efficient in-switch
file-system metadata caching. 

\para{Path-aware cache management (\S\ref{sec:pathaware}).} \sysname caches file-system metadata in the switch data plane by hashing variable-length paths into fixed-length keys and managing file-system metadata in key-value pairs as in existing in-switch key-value caches \cite{li16,jin17,liu19,sheng25}.  Unlike in-switch key-value caches, \sysname addresses path dependencies across hierarchical levels during cache admission and eviction.

\para{Multi-level read-write locking (\S\ref{sec:lock}).}  To synchronize concurrent metadata operations, \sysname adopts multi-level read-write locking using (i) lock counters as read-write locks for different path levels and (ii) validation flags for verifying the validity of cached paths.  It ensures reliable locking and unlocking during path resolution, even in unreliable
networks.

\para{Local hash collision resolution (\S\ref{sec:token}).} Mapping paths to
fixed-size keys in the switch data plane can cause hash collisions, leading to
incorrect metadata retrieval.  For example, if two paths \texttt{/a} and
\texttt{/b} share the same hash and \texttt{/a} is cached, a request for
\texttt{/b} can erroneously return \texttt{/a}'s metadata. \sysname proposes a
{\em local} hash collision resolution approach, where the controller assigns
unique values, called {\em tokens}, to hash-colliding paths and distributes
the tokens to clients, servers, and the switch, so as to allow local
resolution of hash collisions without the controller's intervention for every
request. \sysname ensures no incorrect metadata retrieval.

\subsection{Deployment Modes}\label{subsec:modes}

\sysname supports two deployment modes that share an identical switch-side design: for a read request, the switch performs path resolution against its in-switch cache and serves the request directly on a cache hit; on a cache miss, it forwards the request to the corresponding server. All write requests are forwarded under write-through caching (\S\ref{subsec:goals}). The two modes differ only in client-side configuration and server-side processing.

The \emph{standalone} mode targets deployments where client-side caching is undesirable (e.g., under frequent client churn or large client populations). Clients have no local metadata cache and forward requests directly to the switch; the server processes forwarded requests as in vanilla distributed file systems.

The \emph{complementary} mode targets deployments where client-side caching is already in place but cannot resolve server-side load imbalance under skewed access. Clients additionally cache directory permission metadata locally \cite{ren14,lv22}, resolve directory permissions from their local cache, and forward each request to the switch together with the local permission check results and the versions of the cached permissions used. The server processes the request based on its own authoritative state; if the attached permission versions are stale or absent, the server additionally returns the latest permission metadata for the client to refresh its local cache. We evaluate \sysname (standalone) and \sysnameplus (complementary) in \S\ref{sec:evaluation}.

\section{Path-Aware Cache Management}
\label{sec:pathaware}

\subsection{Path Representation}
\label{subsec:pathhashing}

\sysname incorporates path awareness into cache admission and eviction
by caching frequently accessed paths and their ancestors, so as to mitigate
cache misses during path resolution.  This ensures that if a path is cached,
its ancestors are also cached, thereby allowing the reuse of cached metadata
across different paths with the common ancestors. 

\sysname maps a file-system pathname into fixed-size hash keys for efficient
switch operations. For a read operation, a \sysname client partitions a path
into multiple levels (including the root) and computes a hash for each level.
For example, \texttt{/a/b/c.txt} is partitioned into \texttt{/}, \texttt{/a},
\texttt{/a/b}, and \texttt{/a/b/c.txt}, with each level being hashed.  To
avoid redundant computations, the hash of the root directory \texttt{/} is
pre-computed and cached on the client side.  For a write operation, the client
computes a hash for the complete path without partitioning, as all writes are
forwarded to the server (that holds the namespace) for path resolution under
write-through caching (\S\ref{subsec:goals}).  Currently, \sysname uses the
first 64~bits of a 128-bit MD5 hash as the hash value, where MD5 is used
for fast hashing. Collision resistance is addressed
separately via tokens (\S\ref{sec:token}).

\sysname caches file-system metadata as key-value records in the switch.
Each record is identified by its hash key, and its value contains file or
directory metadata.  Files and directories have common metadata fields (i.e.,
type, permissions, owner/group IDs, and timestamps); while files additionally
include the size and replication factor fields.  Each file's metadata has
40~bytes, and each directory's has 24~bytes.

\sysname manages cache across multiple stages in the switch. It leverages MATs across two stages in a single ingress pipeline to perform cache lookups using hash keys, thereby avoiding cross-pipeline synchronization (\S\ref{sec:lock}) and enforcing logical dependency (i.e., cache hit/miss status is resolved before metadata retrieval can proceed).  Since packets are strictly processed in ingress-to-egress order, this placement enables fast cache status checks. As essential cache status checking and other pipeline operations consume all available stages in the ingress pipeline, the switch allocates 32 register arrays across eight egress stages to store cached metadata values (\S\ref{sec:impl}). To simplify routing, the switch caches metadata in the egress pipeline that is connected to the corresponding server. 

Clients issue metadata requests, including hash keys and full pathnames, to the switch, which reports frequently accessed pathnames to the controller during cache admission (\S\ref{subsec:admission}).  For reads, the switch performs path resolution by issuing cache lookups and permission checks for each level, starting from the root, and recirculates the request to the ingress pipeline for each next level. Due to limited stages and memory access constraints (\S\ref{subsec:switch}), path resolution across multiple levels cannot be completed in a single pass but instead requires one recirculation per level; the corresponding overhead is quantified in \S\ref{subsec:overall-performance}.
For a cache hit (i.e., the metadata for the path and all its ancestors are cached), the switch returns the metadata if all levels' permission checks pass, or an error if any check fails.
To serve cache hits directly from the egress pipeline without an additional egress-to-ingress recirculation round-trip, \sysname uses {\em packet cloning}: the switch clones the in-flight request, embeds the retrieved metadata into the clone, rewrites the clone's destination to the originating client, and drops the original packet. This allows the traffic manager to forward the response immediately.
For a cache miss, the switch forwards the request to the corresponding server. The switch also reports frequently accessed pathnames to the controller to drive cache admission decisions (\S\ref{subsec:admission}). Since writes require multi-level locking to ensure consistency, their processing differs fundamentally from reads; the detailed write workflow is deferred to \S\ref{sec:lock}.

\subsection{Cache Admission and Eviction Workflows}
\label{subsec:admission}

\noindent
{\bf Data structures.} \sysname monitors path access frequencies for read
operations (excluding access to ancestors during path resolution).  For
uncached paths, which dominate path access traffic, \sysname uses a Count-Min
Sketch (CMS) \cite{cormode05}, as in prior studies \cite{jin17,liu19,sheng25},
to estimate their access frequencies within fixed-size memory with provable
error bounds.  For cached paths, \sysname tracks their exact access
frequencies using a frequency counter array.  The switch periodically reports
the access frequencies of all cached paths to the controller for eviction
decisions, and resets the CMS and frequency counter array after each
reporting.

\para{Cache admission.}  \sysname's cache admission is triggered by the switch
data plane. During runtime, the switch monitors the access frequencies of
uncached paths via the CMS and identifies {\em hot} paths (i.e., those
exceeding a pre-defined CMS threshold) for admission. When a hot path $p$ is
detected, the switch notifies the controller, which retrieves the metadata for
$p$ and its uncached ancestors from the servers. The controller communicates
with servers using UDP (and issues retransmissions if needed), bypassing the
switch data plane to avoid critical-path overhead, as in prior work
\cite{jin17,sheng25}.

The controller verifies cache capacity.  If the cache is not full, the controller sends the hash keys and metadata for $p$ and its uncached ancestors to the switch for admission (note that the controller maintains a global view of all cached paths). Unlike NetCache \cite{jin17} (without path awareness), \sysname admits metadata for both $p$ and its uncached ancestors to ensure path-aware caching. To ensure cache consistency during cache admission, while the controller is adding a path to the cache, write requests to this path are blocked at the server until the server is notified that cache admission is finished \cite{jin17}.  Since file-system metadata workloads are read-intensive (\S\ref{subsec:goals}), the blocking action has limited performance impact.

\para{Cache eviction.} When the cache is full and an uncached hot path $p$ is reported, the controller triggers cache eviction to reclaim cache space for $p$ and its uncached ancestors.  It selects cached path candidates for eviction based on periodically reported access frequencies, prioritizing the least frequently accessed path with no cached descendants.  If the selected path is the only cached child of its parent, the parent is also selected.  The controller recursively includes ancestors as candidates until reaching an ancestor with multiple cached children or the root.  The rationale of this recursive inclusion is that the selected ancestors are unlikely to be accessed by themselves, as directory operations are infrequent in practice \cite{ren14,lv22,xu24} (e.g., only 4.2\% of metadata operations in Alibaba's Pangu file system are
directory-related \cite{lv22}).
By reclaiming ancestral slots only when they no longer support any cached descendants, \sysname ensures path resolution integrity and maximizes SRAM utilization.

The controller selects candidates up to a pre-defined threshold, currently set as 2$\times$ the number of paths to be admitted (i.e., $p$ and its uncached ancestors).  A larger threshold reduces the risk of accidentally evicting recently active paths, since the controller's view of access frequencies may be slightly stale between reporting intervals; however, it increases the number of control-plane read packets issued to the switch. In practice, \sysname is robust to this parameter in both directions. A threshold slightly below 2$\times$ has a negligible impact on throughput: any mistakenly evicted path is rapidly re-admitted in the subsequent reporting interval (as seen from Exp\#8 in \S\ref{subsec:microbenchmark} under dynamic workloads); a threshold above 2$\times$ incurs only marginal additional overhead: for a path depth of nine (our evaluation default; \S\ref{subsec:methodology}), even a 3$\times$ threshold introduces at most nine additional read packets over the 2$\times$ default, negligible compared to the round-trip latency of cross-node synchronization and metadata retrieval incurred per admission. Thus, \sysname adopts 2$\times$ as the default.

The controller reloads the current access frequencies of the selected candidates from the switch and evicts the least frequently accessed path with no cached descendants, along with any ancestors having only one cached child. It repeats the selection until sufficient cache space is reclaimed. Finally, the controller notifies the switch to evict the selected paths and admits $p$ and its uncached ancestors. 

\begin{figure}[!t]
\centering
\includegraphics[width=0.9\linewidth]{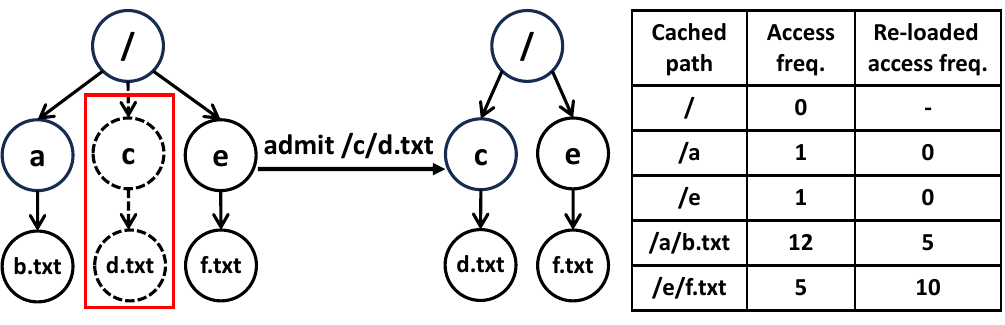}
\caption{Example of cache admission and eviction workflows.}
\label{fig:cachemanagement}
\vspace{-6pt}
\end{figure}

\para{Example.} Figure~\ref{fig:cachemanagement} depicts the cache admission
and eviction workflows. Consider a full cache with five records, holding
\texttt{/}, \texttt{/a}, \texttt{/e}, \texttt{/a/b.txt}, and \texttt{/e/f.txt}
with access frequencies 0, 1, 1, 12, and 5, respectively, where \texttt{/} is
always cached.  Suppose that the switch reports an uncached hot path
\texttt{/c/d.txt}, with a frequency 10, to the controller.  As
\texttt{/c} is uncached, the controller aims to admit both \texttt{/c} and
\texttt{/c/d.txt}.  With a full cache, the controller triggers cache eviction
to select four candidates (i.e., \texttt{/a}, \texttt{/e}, \texttt{/a/b.txt},
and \texttt{/e/f.txt}), twice the number of paths to be
admitted, and reloads their current frequencies (0, 0, 5, and 10,
respectively).  The path \texttt{/a/b.txt} has the lowest access frequency
among all paths without cached descendants, and its ancestor \texttt{/a} (with
only one cached child) will also be evicted.  The controller notifies the
switch to evict \texttt{/a} and \texttt{/a/b.txt}, and admits \texttt{/c} and
\texttt{/c/d.txt}. 

\section{In-Switch Read-Write Locking}
\label{sec:lock}

\subsection{Lock Design}
\label{subsec:lockcounters}

\sysname adopts multi-level read-write locking to ensure cache consistency
under concurrent metadata updates.  The locking mechanism operates entirely
within the switch data plane to avoid controller overhead.  It employs multiple {\em lock counter arrays} and a {\em validation array}, both implemented as register arrays in the switch data plane and initialized with zero entries.  Each cached path is associated with one lock counter (a slot in a lock counter array) and one validation flag (a slot in the validation array), indexed by the path's hash key. 

\para{Lock counter arrays.} \sysname assigns one lock counter array to each of eight path-level groups: the first seven arrays serve levels 1 through 7 individually (e.g., the level-1 array covers paths of the form \texttt{/a}; \S\ref{subsec:metadata}), and the eighth array serves all paths at depth 8 and beyond, keyed by their level-8 hash component. Each array contains 65,536 counters, indexed by the last 16 bits of a path's hash key; each counter records the number of active read requests for the paths mapped to that slot. A cached path is associated with its array by level and with a specific counter by the last 16 bits of its hash key. Concentrating dedicated arrays on the shallowest levels is justified by workload evidence: 90\% of accessed paths have a depth of no more than 10 \cite{agrawal07,meyer12}, with some workloads approaching 100\% \cite{douceur99}. With 65,536 counters per array, the probability that two distinct paths at the same level share a slot is 1/65,536, keeping both slot collisions and lock contention low. Each counter is 16~bits wide, supporting up to 65,535 concurrent read requests per slot to sustain high read parallelism. In case of a slot collision (i.e., two or more paths share the same last-16-bit index at the same level), the counter tracks the aggregate read count across all colliding paths, implying that a write to any one of the colliding paths must wait until all reads at that slot complete; however, the low slot-collision probability makes this overhead negligible in practice.

\para{Validation array.}  The validation array \cite{jin17,sheng25} tracks
metadata validity for all cached paths. A validation flag of one indicates
valid metadata and allows reads from the cache, while a flag of zero indicates
invalid or updating metadata and all reads are directed to servers. After a
cache update, its validation flag is set to one to permit subsequent reads. 

\subsection{Read and Write Flows}
\label{subsec:lockflow}

\sysname places the validation array across all egress pipelines, co-located 
with value register arrays (\S\ref{subsec:pathhashing}) for efficient
validity checks.  Lock counter arrays reside in a single ingress pipeline to
avoid cross-pipeline synchronization.  \sysname redirects requests arriving at
other ingress pipelines to the ingress pipeline holding lock counter arrays
via cross-pipeline recirculation (\S\ref{subsec:methodology}).  This placement is
critical, as lock counter arrays should be positioned before the validation
array to ensure correctness, and they cannot be placed in egress pipelines due
to insufficient switch resources.  In \sysname's deployment, the switch data
plane is not a bottleneck, and consistent packet processing ordering is
maintained across ingress and egress pipelines, so \sysname ensures efficient
and correct locking and validation operations.

\para{Reads.} Reads are classified as (i) {\em single-path reads} (e.g.,
\texttt{stat}), which retrieve only metadata of the requested path;
and (ii) {\em multi-path reads} (e.g., \texttt{readdir}), which retrieve
metadata of the requested path and its descendants.
Single-path reads are served from the in-switch cache, while
multi-path reads are forwarded to servers to ensure correctness, as descendant
paths may be uncached and servers maintain the authoritative namespace to
resolve partially caching scenarios.  Multi-path reads are rare in practice
(e.g., only 3.9\% in Alibaba's workloads \cite{lv22}; see
Table~\ref{tab:workloads}).
Upon receiving a (single-path) read
request, the switch checks if the path's last level is cached, implying that
all its ancestors are also cached (\S\ref{sec:pathaware}). If so, the switch
increments the lock counter for each path level by one and resolves the path
via recirculation.

During path resolution, the switch checks the validation flag for each level.
If the validation flag is one (i.e., valid metadata and no ongoing write), the
switch retrieves the metadata from cache and performs permission checks.  If
the permission checks pass, the switch proceeds to the next level and
decrements the lock counter for the previous level by one.  After resolving
the whole path, the switch decrements the lock counter for the last level by
one via one extra recirculation,
ensuring that all lock counters are released.  If permission checks fail
at any level, the switch sends an error response to the client and decrements
the lock counters from the failure point to the requested path by one.
Conversely, if the validation flag is zero (i.e., invalid or updating
metadata), the switch forwards the read request to the server, which returns a
response.  The switch then decrements all lock counters from the invalid
metadata point to the requested path by one, and returns an ACK to the server.
If the server does not receive the ACK before timeout, it retransmits the same
response. 

\begin{figure}[!t]
\centering
\includegraphics[width=0.8\linewidth]{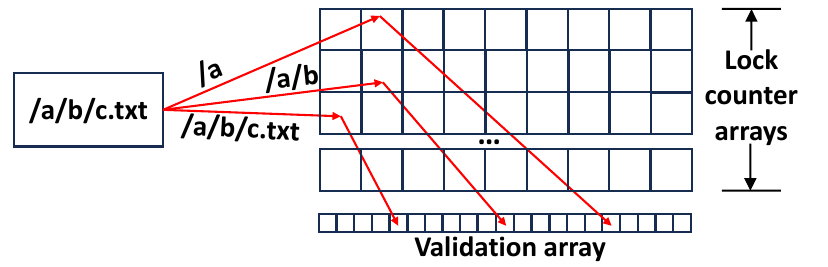}
\caption{Example of processing a read request under multi-level read-write
locking.}
\label{fig:lock}
\vspace{-9pt}
\end{figure}

For example, consider three cached paths \texttt{/a}, \texttt{/a/b}, and \texttt{/a/b/c.txt} (Figure~\ref{fig:lock}). For a read request \texttt{open /a/b/c.txt}, the switch maps each path level \texttt{/a}, \texttt{/a/b}, and \texttt{/a/b/c.txt} to a lock counter in the first, second, and third lock counter arrays, respectively, with corresponding validation flags.  The switch increments the lock counters for all levels by one.  If \texttt{/a}'s validation flag is one and its permission check passes, it processes \texttt{/a/b} and decrements \texttt{/a}'s lock counter by one.  If \texttt{/a/b}'s validation flag is zero, it forwards the request to the server, which retrieves \texttt{/a/b/c.txt}'s metadata. The server returns a response, which triggers the switch to decrement the lock counters for both \texttt{/a/b} and \texttt{/a/b/c.txt} by
one (assuming no packet loss).  The switch then returns an ACK to the server.

\para{Writes.} Writes are classified as (i) {\em single-path writes} (e.g.,
\texttt{chmod} and \texttt{chown}), which update only the requested path, and
(ii) {\em multi-path writes} (e.g., \texttt{chmod -R} and \texttt{chown -R}),
which update the requested path and its descendants. The locking mechanism
ensures consistency for both. 

{\em (i) Single-path writes.}  For a single-path write, the switch checks if
the path is cached.  If so, it checks the corresponding lock counter based on
its level and hash key.  If the lock counter is non-zero (i.e., with ongoing
reads), the switch recirculates the request until the lock counter reaches
zero (i.e., no ongoing read).  Then, the switch sets the validation flag to
zero and forwards the request to the server.  If the write is successfully
completed, the server returns a response, which triggers the switch to update
the cached path's metadata and set the validation flag to one; otherwise, only
the validation flag is set to one without cache updates. 

The current reader-preferring lock design is susceptible to writer starvation: while a write request recirculates waiting for the lock counter to reach zero, new incoming reads increment the counter before the write can acquire the lock. Under a continuous stream of read requests, the counter never reaches zero, blocking the write indefinitely. A potential remedy is to block new reads once a pending write is detected at the switch (e.g., by setting a write-pending flag that arriving reads must check before incrementing the counter \cite{zhang24}). We acknowledge writer starvation as a current limitation and leave the design of a starvation-free locking mechanism as future work.

{\em (ii) Multi-path writes.}  Multi-path writes follow the same single-path
write processing until the request reaches the server.  If the write is
successfully completed, the server updates the cache for all cached
descendants before the requested path, so that the requested path remains
invalidated until all cached descendants are fully updated.  By performing
path resolution in a top-down manner, \sysname prevents reads from accessing
cached descendants before the requested path is updated (i.e., until the cache
updates for a multi-path write are completed) to ensure cache consistency. 

\section{Local Hash Collision Resolution}
\label{sec:token}

\begin{figure}[!t]
\centering
\includegraphics[width=0.9\linewidth]{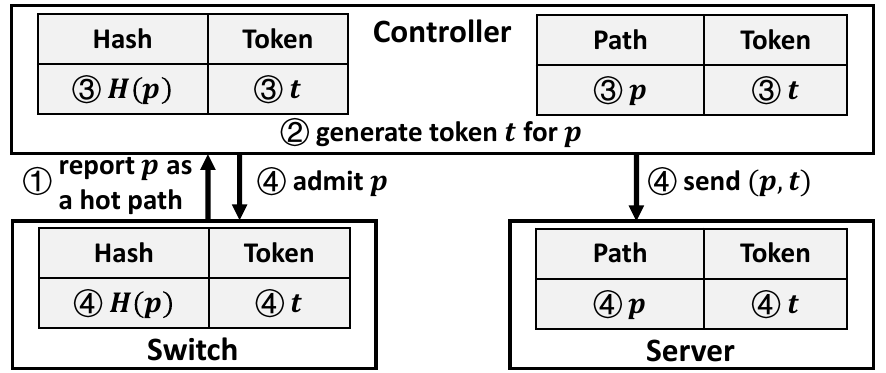}
\caption{Example of token generation and distribution.  Note that the
controller also assigns tokens for $p$'s uncached ancestors in Steps~2-4. We
omit the details for brevity.}
\label{fig:token1}
\vspace{-9pt}
\end{figure}

\sysname maps file-system pathnames to 64-bit hash keys (\S\ref{subsec:pathhashing}), so hash collisions are possible (albeit unlikely) and lead to incorrect metadata retrieval.  While the controller holds a global view of all cached paths (\S\ref{subsec:admission}), querying the controller to resolve hash collisions for every request incurs high switch-to-controller latencies (\S\ref{subsec:challenges}). \sysname adopts a local, token-based hash collision resolution mechanism by synchronizing the controller's global view of cached paths with the switch, clients, and servers, preserving correctness and high performance.

\subsection{Token-Based Hash Collision Resolution}
\label{subsec:tokenhash}

\noindent{\bf Token definition and structures.}
A {\em token} is an 8-bit value
paired with a 64-bit hash key to uniquely identify a path.  {\em Valid} tokens
range from 1 to 255, while 0 indicates {\em invalid}.  A cached path is
assigned token 1 if no collision occurs, or the next available token (e.g., 2)
if it collides with an existing cached path. 
\sysname maintains two unordered map structures: (i) a {\em path-token map},
which records paths and their tokens ({\em path-token pairs}), and (ii) a
{\em hash-token map}, which records hash keys and their tokens ({\em
hash-token pairs}).  The controller keeps both maps, each client and server
holds a path-token map, and the switch holds a hash-token map.
All maps are kept in memory for fast lookup. For failure recovery (\S\ref{subsec:recovery}), the controller additionally maintains two persistent logs in a local key-value store (e.g., RocksDB~\cite{rocksdb}), each recording a snapshot of both maps: (i) an {\em active log} covering paths currently in the cache, used to repopulate the switch cache and hash-token map after a switch failure and to restore a recovered server's path-token map; and (ii) a {\em historical log} covering all paths ever cached, used to recover the controller's own maps upon restart.

\para{Token generation.} During cache admission, the controller assigns tokens to each level of a hot path, as shown in Figure~\ref{fig:token1}.  When the switch reports a hot path $p$ for admission (\circled{1} in Figure~\ref{fig:token1}), the controller checks its map structures. If $p$ is the first time being admitted and its hash key is unique, the controller assigns a token of value one; if a collision occurs, it assigns the next available token (e.g., $t$)
(\circled{2} in Figure~\ref{fig:token1}).  The controller updates its path-token and hash-token maps with the new path-token and hash-token pairs, respectively (\circled{3} in Figure~\ref{fig:token1}).
These entries persist in the controller's maps even after eviction: if an evicted path is later re-admitted, its previously assigned token is reused, ensuring consistent token identity across admission cycles. To support failure recovery, cache admission appends the new entry to both the active and historical logs, while cache eviction removes the entry only from the active log, with the historical record preserved.

\para{Token distribution.}
During cache admission, the controller distributes each admitted path and its token to the switch and
the relevant server that holds the path (\circled{4} in Figure~\ref{fig:token1}).  The switch adds the
hash key and token to the hash-token map, while the server adds the full
path and token to the path-token map.  These entries are removed during cache
eviction, as notified by the controller.

\begin{figure}[!t]
\centering
\includegraphics[width=0.99\linewidth]{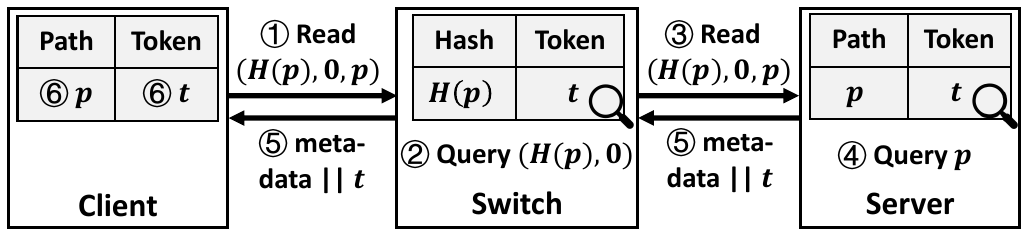}
\caption{Example of how a client updates its path-token map. Note that
the client also attaches tokens of $p$'s ancestors in Step~1, the server
returns the tokens for $p$'s ancestors in Step~5, and the client adds the
tokens in Step~6. We omit the details for brevity.}
\label{fig:token2}
\vspace{-9pt}
\end{figure}

\para{Token discovery and utilization.} Clients update their path-token maps when
issuing read or write requests, as shown in Figure~\ref{fig:token2}.
Initially, with an empty path-token map, a client attaches an invalid token
(value zero) to a request (e.g., reading path $p$'s metadata) sent to the
switch (\circled{1} in Figure~\ref{fig:token2}).  The switch detects a cache miss by querying its
hash-token map (\circled{2} in Figure~\ref{fig:token2}) and forwards the request to the server
(\circled{3} in Figure~\ref{fig:token2}).  The server retrieves the metadata, checks its path-token map
(\circled{4} in Figure~\ref{fig:token2}), and responds with the metadata and valid tokens (\circled{5} in Figure~\ref{fig:token2}).
The client updates its path-token map with the valid tokens for future
requests (\circled{6} in Figure~\ref{fig:token2}).  For cached paths, the switch confirms cache hits when
valid tokens are included in the request.  

\subsection{Token Scalability and Overhead Analysis}
\label{subsec:tokenanalysis}

\noindent{\bf Scalability analysis}. \sysname's scalability depends on two properties: its ability to tolerate hash collisions among cached paths, and its ability to store mapping entries across a growing namespace without unbounded resource consumption.

For collision tolerance, \sysname uses 8-bit tokens to distinguish up to 255 distinct paths that map to the same 64-bit hash index. For 64-bit hash keys, the pairwise collision probability between any two specific paths is $1/2^{64}$. Given a cache of $C$ entries, the expected number of cached paths sharing any particular hash index is $C/2^{64}$, which is close to zero even for billion-scale namespaces. The probability that any single hash index accumulates more than 255 cached paths is hence extremely small, and 8-bit tokens provide sufficient collision capacity for any practical file-system deployment.

For storage scalability, the controller's maps grow monotonically: since token assignments persist in the controller's maps even after cache eviction (\S\ref{subsec:admission}), re-admitted paths can reuse their previously assigned tokens. In the worst case, the controller accumulates entries for all distinct paths that have ever been cached.
Since the controller persists its maps to a local key-value store for failure recovery (\S\ref{subsec:tokenhash}), the maps can scale beyond DRAM capacity by leveraging the key-value store's on-disk storage.
In contrast, storage requirements for the switch and servers are strictly bounded by the current cache size, since entries are installed upon admission and removed upon eviction, so the active state in the switch data plane and on each server scales with the cache size rather than the total namespace. Client storage is bounded by the active working set: clients attach an expiry time (e.g., one hour) to each local path-token entry and periodically clear stale entries, preventing indefinite accumulation.

\para{Overhead analysis.} Fletch's token mechanism incurs minimal communication and storage overhead. For each request, a client attaches one (hash, token) pair per path level (including the root), contributing 9 bytes per level (an 8-byte hash key and a 1-byte token). The total per-request PHV overhead for a path of depth $d$ is $9(d+1)$~bytes, much smaller than the native HDFS representation, where each path component can reach 255 bytes (\S\ref{subsec:challenges}). This compact per-level encoding substantially alleviates pressure on the switch's limited PHV capacity (768~bytes on Tofino). Although per-request PHV overhead grows linearly with path depth, it remains practically small because real-world workloads exhibit small path depths (\S\ref{subsec:lockcounters}; for example, 90\% of accessed paths have a depth of no more than 10).

The storage overhead of the token mechanism is small across all components. In the switch, the hash-token map stores 9-byte entries (an 8-byte hash key and a 1-byte token); it is implemented using MATs rather than register arrays to enable efficient exact-match lookups without the entry-size restrictions of SRAM registers. The controller maintains a hash-token map covering all paths ever cached rather than only currently cached ones, alongside its path-token map; the path-token map is also held by servers and clients. For these path-token maps, each entry pairs the full path string with a 1-byte token. Since path strings are inherent to the file-system namespace and are always stored, the extra storage cost incurred by the token mechanism is at most 1 byte per cached path across all components.

\section{Failure Analysis}
\label{sec:failure}

We present \sysname's complete failure-handling design across clients, the switch, the controller, and servers, covering cache admission (\S\ref{subsec:admissionfailures}), request serving (\S\ref{subsec:servingfailures}), and recovery (\S\ref{subsec:recovery}). A few mechanisms are described as extensions beyond our current prototype (\S\ref{sec:impl}).

\subsection{Failures during Cache Admission}
\label{subsec:admissionfailures}

Cache admission is orchestrated among the switch, controller, and servers (\S\ref{subsec:admission}), so \sysname must remain resilient to both packet loss and component failures during this process. Note that server failures are orthogonal to cache admission correctness: since the cache state and management logic reside exclusively in the controller and switch, a server failure does not corrupt the global cache state. Server-level fault tolerance during cache admission is delegated to the backend distributed file system, which provides replication and recovery, ensuring that the controller can always retrieve metadata for a path under admission from a surviving replica.

\para{Packet loss.} \sysname ensures end-to-end reliability for the admission protocol through proactive redundancy and ACK-based retransmission. Communication between the controller and servers uses UDP with a timeout-and-retry mechanism to guarantee reliable delivery of both metadata fetches and cache-completion notifications. On the switch-to-controller channel, the switch proactively transmits each hot-path report and admission acknowledgment multiple times (three by default) to compensate for data-plane packet loss \cite{jin17}. On the controller-to-switch channel, the controller reissues a cache update packet if no acknowledgment is received within a configured timeout. These mechanisms ensure that the admission handshake always terminates, preventing writes from being indefinitely blocked or leaving paths in an inconsistent intermediate state.

\para{Switch and controller failures.} Recall that write requests to the path undergoing admission are held at the server until the server receives a cache-completion notification from the controller, which is triggered only after the switch acknowledges the cache update (\S\ref{subsec:admission}). Thus, a failure of either the switch or the controller can stall the admission chain, blocking the write indefinitely. To prevent this, \sysname employs a server-side timeout mechanism: if the server does not receive the cache-completion notification within a preconfigured window, it lifts write blocking, registers an admission failure, and resumes normal request handling. This ensures that a failure in either the data plane or the control plane does not permanently impair the backend file system's availability. 

\subsection{Failures during Request Serving}
\label{subsec:servingfailures}

\noindent{\bf Packet loss.} A lost switch-to-server request or server-to-switch response causes the client to time out and retransmit, ensuring eventual delivery. However, a subtle correctness issue arises when a retransmitted read targets a path whose validation flag is currently zero (i.e., being updated by a concurrent write). In this case, the switch increments the affected lock counter without a corresponding decrement from the earlier lost exchange, thereby permanently elevating the counter and blocking subsequent writes to that path. \sysname addresses this by having the controller periodically monitor lock counters and reset stale ones.

The controller distinguishes a stale counter from a legitimate in-flight read by observing that any in-flight read continually drives counter changes: each request, retransmission, or response triggers an increment or decrement. The maximum interval between two consecutive such events is bounded by $T_{\text{event}}$, determined by the client retransmission timeout, network round-trip time, and server processing time (typically hundreds of milliseconds). The controller sets its detection window $T_{\text{stale}} \gg T_{\text{event}}$ (e.g., $T_{\text{stale}} = 30$\,s), so that any counter unchanged for longer than $T_{\text{stale}}$ cannot be referenced by a live request and is safely classified as stale.

Once a stale counter is identified, the controller records its observed value $v_{\text{obs}}$ and embeds it in a control packet sent to the switch. The switch performs an atomic conditional reset \cite{joshi23}: it sets the counter to zero only if its current value still equals $v_{\text{obs}}$. If a new read has incremented the counter between detection and reset, the conditional reset is skipped, and the controller re-evaluates the counter in the next monitoring round.

The switch-to-server ACK loss is particularly critical: the server retransmits its response upon a timeout, causing the switch to apply a duplicate lock decrement, violating system correctness.
\sysname prevents this via a per-server sequence-number protocol.
Each server tags every lock-related response with a monotonically increasing sequence number
(initialized to 0) and increments it only upon receiving an ACK from the switch.
The switch maintains a per-server expected sequence number in a sequence counter array.
Upon receiving a response, the switch compares its embedded sequence number against the expected value.
If they match, the switch processes the lock update, forwards the response to the client,
returns an ACK to the server, and increments the expected value.
If the embedded sequence number is strictly less than the value,
the response is a retransmission; the switch returns an ACK to suppress
further retransmissions without applying any lock update.

\para{Server failure.} Server failures are handled by the backend distributed file system, which provides the necessary replication and fault tolerance to preserve file-system consistency independently of \sysname. As \sysname employs write-through caching, every cached metadata entry is always consistent with the corresponding server-side record, and no additional cache reconciliation is required upon server recovery.

\para{Controller failure.} A controller failure does not affect requests currently in flight, as the switch data plane and backend servers operate independently during request serving. The controller's role in the request path is limited to cache admission and eviction, both of which are non-critical for serving existing cache entries. Any ongoing cache admission is gracefully aborted via the server-side timeout mechanism described in \S\ref{subsec:admissionfailures}, and \sysname continues to serve cached requests from the switch until the controller recovers (\S\ref{subsec:recovery}).

\para{Client failure.} A client crash does not affect the correctness of the in-switch cache or the server-side namespace. Clients hold only local path-token pairs (transient state used for hash-collision resolution) and carry no authoritative cache records. Upon recovery, the client's path-token map is empty. Each subsequent request experiences at most one cache miss per cached path to retrieve the correct token from the server (\S\ref{subsec:tokenhash}), after which normal cache-hit operation resumes.

\para{Network partition}. A partition between the controller and the switch halts new cache admissions and evictions, but requests for already-cached entries are still served by the switch data plane and backend servers. During the partition, the cache content remains unchanged because no admission or eviction can proceed. Once the partition heals, the controller cross-checks its cached-path records against the switch's current register state and reissues any pending admission or eviction operations before resuming normal operations.

\subsection{System Recovery}\label{subsec:recovery}

When the switch, controller, or server experiences a total crash, \sysname restores a globally consistent state through a targeted recovery procedure that minimizes downtime and avoids a full cache cold start whenever possible.

\para{Controller failure.} If only the controller fails, the switch
continues serving cached requests normally during the outage.
Upon controller restart, the controller reconstructs its path-token and hash-token maps from the historical log in its local persistent key-value store (\S\ref{subsec:tokenhash}), which records all
token assignments, including those for previously evicted paths. This ensures that re-admitted paths retain their original tokens, preserving consistency with any path-token pairs cached at clients. Since all token assignments are recovered from persistent storage, the controller restores a complete view of the cached namespace without flushing the switch cache or re-running hot-path identification, and then resumes admission and eviction responsibilities with minimal disruption.

\para{Switch failure.} A switch failure resets all in-switch data-plane state,
including the metadata cache, hash-token map, lock counters, and validation flags.
Upon reboot, \sysname replays cache admission for each
path recorded in the controller's active log (\S\ref{subsec:tokenhash}):
the controller re-fetches the corresponding metadata from the backend
servers and re-installs both the metadata and the hash-token map into
the switch data plane, with each path retaining its original token.
This bypasses the initial hot-path identification phase and
enables rapid cache warm-up, so \sysname can quickly restore near-peak
performance.

\para{Server failure.} Server failures are recovered by the backend
distributed file system. Since \sysname uses write-through caching,
the cached metadata in the switch remains consistent
with the server's last committed state and does not need to be reconciled
upon server recovery. However, the server's path-token map
(\S\ref{subsec:tokenhash}) is in server memory and is lost upon
a failure. To reconstruct it, the recovered server queries the
controller, which extracts the path-token entries assigned to this
server from the active log (\S\ref{subsec:tokenhash}) and sends them
back via UDP with a timeout-and-retry mechanism for reliable delivery.
Once restored, the server resumes serving requests, and any
cache entries assigned to the recovered server remain valid.

\section{Implementation}
\label{sec:impl}

We have implemented a prototype of \sysname on the Tofino switch chipset \cite{tofino}, comprising the controller, in-switch cache, servers, and clients. Tofino switches build on the Protocol-Independent Switch Architecture (PISA) under the RMT paradigm \cite{bosshart13} and are programmed in P4 \cite{bosshart14}. Although our prototype targets Tofino switches, the core techniques (i.e., path-aware cache management (\S\ref{sec:pathaware}), multi-level read-write locking (\S\ref{sec:lock}), and local hash collision resolution (\S\ref{sec:token})) can be ported to any other RMT-based programmable ASIC \cite{broadcomswitch,chole17,huaweiswitch} that supports packet recirculation and register arrays.

\para{Controller.} The controller is implemented in C++ with 2.8\,K LoC. It
leverages APIs provided by the Tofino switch compiler \cite{tofino} to
interact with the switch data plane, including configurations and updates of
MATs and registers, for cache admission and eviction (\S\ref{sec:pathaware})
and token management (\S\ref{sec:token}). It uses RocksDB \cite{rocksdb} (v6.22.1) to persist the active and
historical logs for failure recovery.

\para{In-switch cache.} The in-switch cache is implemented in P4 with 5.1\,K
LoC and compiled for a Tofino switch \cite{tofino,switchthpt}.  It
comprises several components: (i) a hash-token map using MATs for cache
lookups and local hash collision resolution, where each entry is a 9-byte key
(8-byte hash key and 1-byte token); (ii) 32 register arrays of 32-bit slots
for storing cache entries; (iii) a three-row CMS with three register arrays,
each with 64K 16-bit slots; (iv) a frequency counter array with a register
array of 32-bit slots; (v) lock counter arrays with eight register arrays,
each with 64K 16-bit slots; (vi) a validation array with a register array
of 1-bit slots; and (vii) a sequence counter array with a register array of
8-bit slots.

\para{Server implementation.} \sysname builds on Hadoop Distributed File System (HDFS) \cite{shvachko10}, a popularly deployed distributed file system, where a {\em namenode} (i.e., metadata server) manages the namespace and metadata. Each server is implemented in C++ with 3\,K LoC and hosts an HDFS (v3.2) namenode \cite{hdfs}.  We use RBF \cite{hdfsrbf} with the \texttt{HASH\_ALL} policy for namespace and metadata management, which uses consistent hashing to distribute files evenly across all namenodes for load balancing and creates directories on all namenodes.  Each server manages a subset of the metadata namespace, connects to its local namenode via the C++ HDFS client library \texttt{libhdfs3} \cite{libhdfs3}, and serves client requests while maintaining a path-token map for local hash resolution (\S\ref{sec:token}).

\para{Client implementation.} The C++ client driver supports multi-threaded
workload execution, where each thread simulates a logical client.
Each client maintains a path-token map for local hash collision resolution
(\S\ref{sec:token}). It communicates with the switch via UDP-encapsulated
packets customized for metadata requests, with re-transmission support for
reliability.  We integrate the driver with the \texttt{mdtest} \cite{mdtest}
benchmarking tool for HDFS metadata operations. 

\begin{table}[t]
\footnotesize
\centering
\captionof{table}{Implementation status of failure-handling mechanisms.}
\label{tab:impls}
\vspace{-3pt}
\resizebox{\columnwidth}{!}{
\renewcommand\arraystretch{1.0}
\begin{tabular}{|c|c|}
\toprule[1pt]
\textbf{Mechanism (Section)} & \textbf{Status} \\
\bottomrule[0.5pt]
\toprule[0.5pt]
Packet loss handling during admission (\S\ref{subsec:admissionfailures}) & \cmark \\
\midrule[0.5pt]
Server-side admission-stall timeout (\S\ref{subsec:admissionfailures}) & \xmark \\
\midrule[0.5pt]
Sequence-number protocol for switch-to-server ACK loss (\S\ref{subsec:servingfailures}) & \cmark \\
\midrule[0.5pt]
Stale lock counter monitoring (\S\ref{subsec:servingfailures}) & \xmark \\
\midrule[0.5pt]
Network partition reconciliation (\S\ref{subsec:servingfailures}) & \xmark \\
\midrule[0.5pt]
Controller/switch/server recovery (\S\ref{subsec:recovery}) & \cmark \\
\bottomrule[1pt]
\end{tabular}
}
\vspace{-6pt}
\end{table}

\para{Scope of implementation.} Our prototype covers all components described in \S\ref{sec:pathaware}--\S\ref{sec:token}. Table~\ref{tab:impls} summarizes the implementation status of the fault-tolerance mechanisms in \S\ref{sec:failure}. The unimplemented mechanisms are orthogonal to the core caching framework and can be added without altering \sysname's design.

\para{Portability beyond programmable switches.} \sysname is prototyped on a Tofino P4-PISA switch, yet its core techniques are not inherently tied to the Tofino architecture and are portable to other programmable hardware platforms. One possible future direction is a Data Processing Unit (DPU), which sits at the host edge between the NIC and the host CPU.
In a DPU-based deployment, \sysname would transition from a network-core cache to a host-edge accelerator, eliminating the kernel file-system stack traversal and RPC round-trip overhead for metadata requests, since the DPU intercepts them before they reach the host. A complete mapping of \sysname's core mechanisms (e.g., path resolution, lock counter placement, and multi-DPU coordination for cache consistency) onto DPU hardware requires a non-trivial design and is beyond the scope of this paper.

\section{Evaluation}
\label{sec:evaluation}

\subsection{Methodology}
\label{subsec:methodology}

\noindent
{\bf Testbed.} We evaluate \sysname on a testbed comprising a 3.2\,Tbps
two-pipeline Tofino switch \cite{tofino,switchthpt} and three physical
machines.  Two machines host servers, each with a 2.40\,GHz 10-core Intel Xeon
Silver 4210R CPU, 128\,GiB DRAM, and a 2\,TB HDD (Dell PERC H330 Mini).  The
client driver runs on the remaining machine, with a 2.40\,GHz, 16-core Intel
Xeon Silver 4314 CPU, 128\,GiB DRAM, and a 960\,GB NVMe SSD (Micron 9300 PRO).
Each machine is connected to the switch via a 40\,Gbps NIC (Mellanox
ConnectX-5 CX516A).  The client machine uses one pipeline, and the two server
machines use another pipeline. All counter arrays (\S\ref{sec:impl}) reside
in the server-connected pipeline.

Since our two-pipeline Tofino switch lacks native support for cross-pipeline recirculation, we physically connect the designated ingress pipeline hosting the lock counter arrays to another ingress pipeline using a physical wire \cite{sheng25}. This enables requests from another ingress pipeline to be recirculated to the designated ingress pipeline (\S\ref{sec:lock}). As a result, every client request incurs one cross-pipeline recirculation to access the lock counter arrays in the server-connected pipeline.  Future programmable switches with native cross-pipeline recirculation would eliminate this requirement. 

\para{Workloads.} We evaluate \sysname using four real-world workload traces:
(i) Alibaba's Pangu file-system instances (Alibaba) \cite{lv22}, (ii)
convolutional neural network training (Training) \cite{xu24}, (iii) the
processing of one million thumbnail images (Thumb) \cite{xu24}, and (iv) a
LinkedIn HDFS cluster (LinkedIn) \cite{ren14}.  Table~\ref{tab:workloads}
summarizes the proportion of metadata operations and the read ratio for each
workload.  

\begin{table}[t]
\small
\centering
\captionof{table}{Real-world workloads and their metadata operation ratios.}
\label{tab:workloads}
\vspace{-3pt}
\resizebox{\columnwidth}{!}{
\renewcommand\arraystretch{1.0}
\begin{tabular}{|c|c|c|}
\toprule[1pt]
\textbf{Workload} & \textbf{Operation ratio} & \textbf{Read ratio} \\
\bottomrule[0.5pt]
\toprule[0.5pt]
\multirow{3}*{Alibaba \cite{lv22}} & 52.6\%~open/close,
	9.59\%~create, 3.9\%~readdir, & \multirow{3}*{69.1\%}\\
~ & 0.1\%~chmod, 11.9\%~delete, 12.4\%~stat, 0.2\%~statdir, & ~ \\
~ & 0.005\%~mkdir, 0.005\%~rmdir, 9.3\%~file rename& ~ \\
\midrule[0.5pt]
\multirow{3}*{Training \cite{xu24}} & 54.3\%~open/close, 27.16\%~stat,
0.13\%~readdir, & \multirow{3}*{81.7\%} \\
~ & 9.01\%~create, 0.13\%~mkdir, 0.13\%~rmdir, & ~ \\
~ & 9.01\%~delete, 0.13\%~statdir & ~ \\
\midrule[0.5pt]
\multirow{2}*{Thumb \cite{xu24}} & 57.01\%~open/close, 28.44\%~stat,
0.13\%~readdir, & \multirow{2}*{85.7\%}  \\
~ & 14.16\%~create, 0.13\%~mkdir, 0.13\%~statdir & ~ \\
\midrule[0.5pt]
\multirow{2}*{LinkedIn \cite{ren14}} & 84\% open/getattr, 9\% create/mkdir,&
\multirow{2}*{84\%}  \\
~ & 7\% chmod/delete/rename & ~ \\
\bottomrule[1pt]
\end{tabular}
}
\vspace{-6pt}
\end{table}

We refine the workloads for our evaluation.  To focus on metadata performance,
we exclude reads and writes of file data, following prior studies
\cite{niazi17,li17,ren14,lv22}. For Training and Thumb that include file reads
and writes \cite{xu24}, we exclude these operations and normalize the
remaining metadata operations.  Since we exclude file writes, we treat
\texttt{close} as a read operation, while it updates both modification and
access timestamps if the closed file has been updated.  HDFS updates access
timestamps hourly by default \cite{defaultsetting}, and we exclude timestamp
updates as they are infrequent. 

For LinkedIn, as the original paper \cite{ren14} does not provide operation ratios, we derive them from trace analysis \cite{lv22}: \texttt{open} (42\%), \texttt{getattr} (42\%), \texttt{create} (4.5\%), \texttt{mkdir} (4.5\%), \texttt{chmod} (1\%), \texttt{delete} (3\%), and \texttt{rename} (3\%). We assign the smallest ratio to \texttt{chmod}, which is less frequent than \texttt{delete} and \texttt{rename} \cite{lv22}.  We replace \texttt{getattr} (which corresponds to \texttt{stat} and \texttt{statdir} in HDFS) with \texttt{stat}, as file operations are much more frequent than directory operations \cite{lv22}, and focus on file renaming for \texttt{rename}.

\para{Generation of metadata operations.} We use \texttt{mdtest} \cite{mdtest} to generate file-system namespaces and metadata operations. We configure a default path depth of nine, as metadata requests are often aggregated at small depths \cite{agrawal07,meyer12,douceur99}. We create 32~million empty files to focus on metadata performance, following prior evaluation \cite{ren14,xiao15,niazi17,li17,lv22,wang23}. To simulate workload skewness, we apply the 80/20 rule (80\% of operations on 20\% of directories and files \cite{wang21,xu24}) and model file access frequencies across levels using a power-law distribution with an exponent of 0.9, with operation types assigned randomly using the ratios in Table~\ref{tab:workloads}. For \texttt{statdir} and \texttt{readdir}, we select the parent directory of the chosen file; for \texttt{mkdir} and \texttt{rmdir}, we use separate directories to avoid removing non-empty directories. We issue 32~million metadata requests per workload and address various workload settings in \S\ref{subsec:microbenchmark}.

We mix all metadata operations in each workload.  However, \texttt{rename}, \texttt{delete}, and \texttt{rmdir} involve metadata modifications that necessitate the granting and revoking of leases in HDFS \cite{wang09}, and the lease-based operations on frequently created and deleted files can slow down all metadata operations under stress tests.  Thus, we place them
at the end of the request sequence to keep metadata operations at high rate. 

\para{Scaling servers.} Although our testbed contains only two physical
servers, our Tofino switch has significantly higher forwarding throughput
(3.2\,Tbps \cite{switchthpt}) than server throughput (tens of KOPS) and is
not our evaluation bottleneck.  To simulate larger-scale deployment with
limited hardware while maintaining realism in evaluation, we adopt the
well-known {\em server rotation} approach as in prior in-switch caching
studies \cite{jin17,sheng25}.  Let $S$ be the number of simulated servers.  We
assign files to these $S$ servers using HDFS's RBF policy
(\S\ref{sec:impl}).  We calculate each server's load from file access
frequencies and identify the bottlenecked server with the highest load.  We
then perform our evaluation in $S$ iterations.  We first deploy the
bottlenecked server in one physical machine and fully saturate it to measure
its performance.  In the following $S-1$ iterations, we reset the physical
machines to the initial state, pair the bottlenecked server with one of the
$S-1$ non-bottlenecked servers, deploy them on the two physical machines, and
measure the performance of the non-bottlenecked server.  We aggregate the
performance of all servers as the overall performance.   By default, we
simulate 16 servers and increase the scale to 128~servers (Exp\#1). 

\para{Baselines.}  We consider {\em client-side caching (CCache)}
by faithfully following the state-of-the-art client-side caching
implementations in IndexFS \cite{ren14} and InfiniFS \cite{lv22}
(\S\ref{sec:related}).  CCache uses RocksDB (v6.22.1)
as a key-value store to keep all metadata instead of in an HDFS namenode in
each simulated server to eliminate HDFS's path resolution overhead.  Each
CCache client locally caches directories' permission metadata, and forwards
all read requests to servers for attribute retrieval.  We implement lazy
invalidation \cite{lv22} for cache consistency, which outperforms lease-based
cache management \cite{ren14}.  We do not compare \sysname with in-switch
key-value cache systems (e.g., NetCache \cite{jin17} and FarReach \cite{sheng25})
since they are designed for key-value stores and
cannot ensure correctness or consistency for file-system operations.

We evaluate four schemes organized by deployment mode (\S\ref{subsec:modes}). In the standalone mode: (i) {\em NoCache}, which performs metadata operations directly with HDFS namenodes without caching; and (ii) {\em \sysname}, which extends NoCache with in-switch caching. In the complementary mode: (iii) {\em CCache}, our client-side caching implementation; and (iv) {\em \sysnameplus}, which extends CCache with in-switch caching.

Before each experiment, we pre-load 32 million files into HDFS namenodes for
NoCache and \sysname, and the corresponding paths and metadata into RocksDB
for CCache and \sysnameplus.  We pre-load the 5,000 hottest files and their
ancestors into the in-switch cache for \sysname and \sysnameplus. We set
the pre-defined CMS threshold of \sysname and \sysnameplus to 10 and 20,
respectively (a larger threshold is used for \sysnameplus due to its higher
throughput with client-side caching), and reset the CMS and the frequency
counter array every two seconds. We simulate 128
client threads, which sufficiently saturate backend servers.  The client-side
cache of each simulated client in CCache and \sysnameplus is allocated 4\,MiB
\cite{lv22}.
We plot the average results over five runs, with error bars as 95\%
confidence intervals under the Student's t-distribution.

\subsection{Performance Analysis}
\label{subsec:overall-performance}

\begin{figure}[!t]
\centering
\includegraphics[height=12pt]{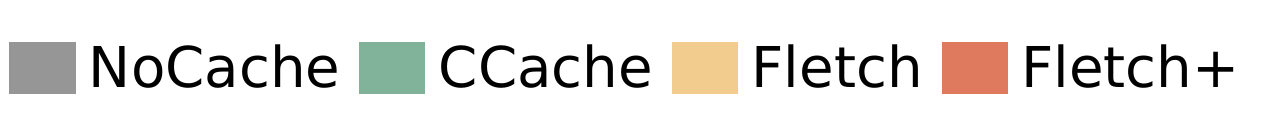}
\begin{tabular}{@{\ }c@{\ }c}
\includegraphics[width=0.495\linewidth]{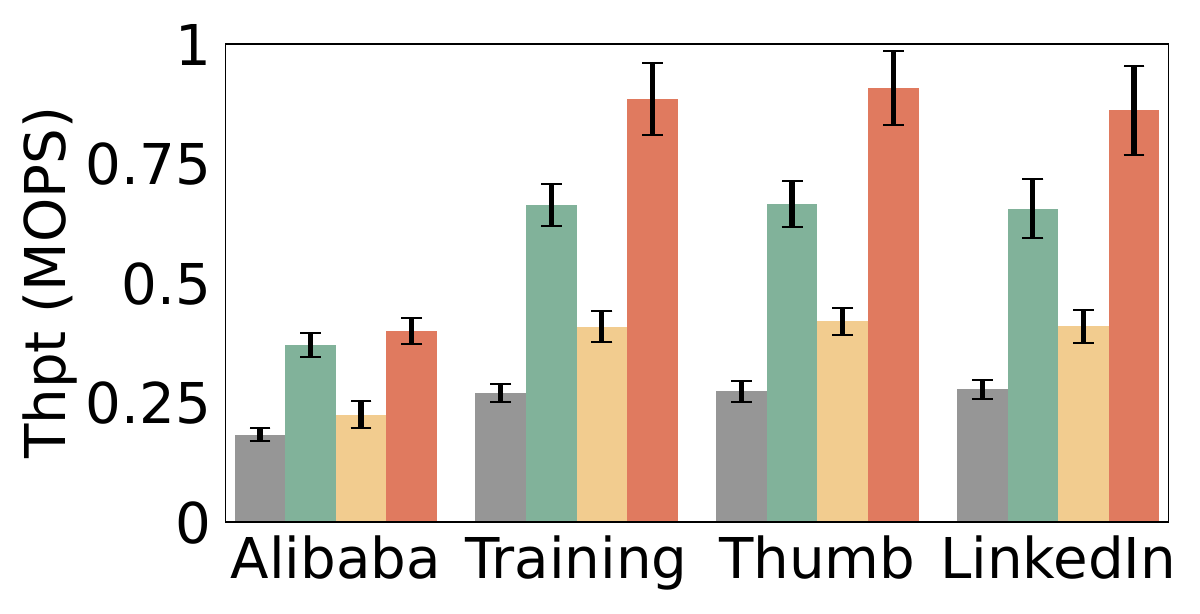}   &
\includegraphics[width=0.495\linewidth]{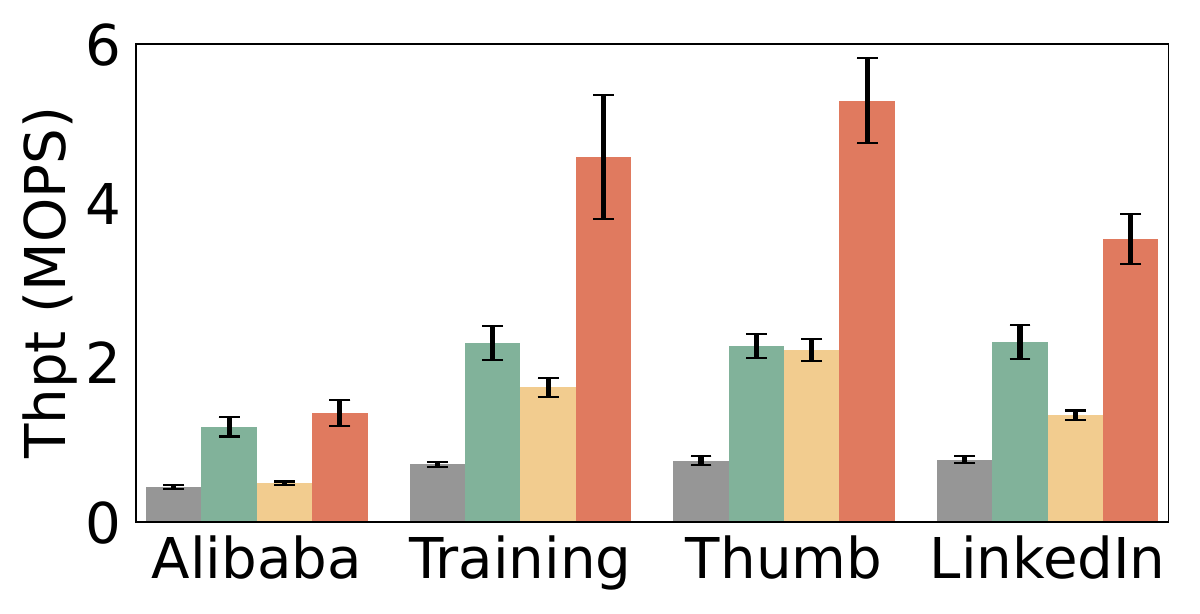}
\vspace{-3pt}\\
\makecell[c]{\small (a) 16 simulated servers} &
\makecell[c]{\small (b) 128 simulated servers}
\end{tabular}
\vspace{-3pt}
\captionof{figure}{(Exp\#1) Performance under real-world workloads.}
\label{fig:exp1}
\vspace{-6pt}
\end{figure}

\begin{figure}[!t]
\centering
\includegraphics[height=14pt]{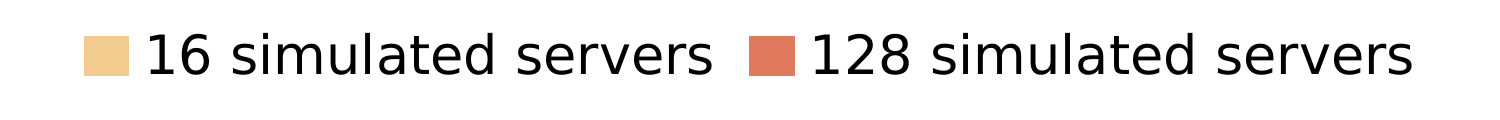}
\vspace{-3pt}\\
\begin{tabular}{@{\ }c@{\ }c}
\includegraphics[width=0.49\linewidth]{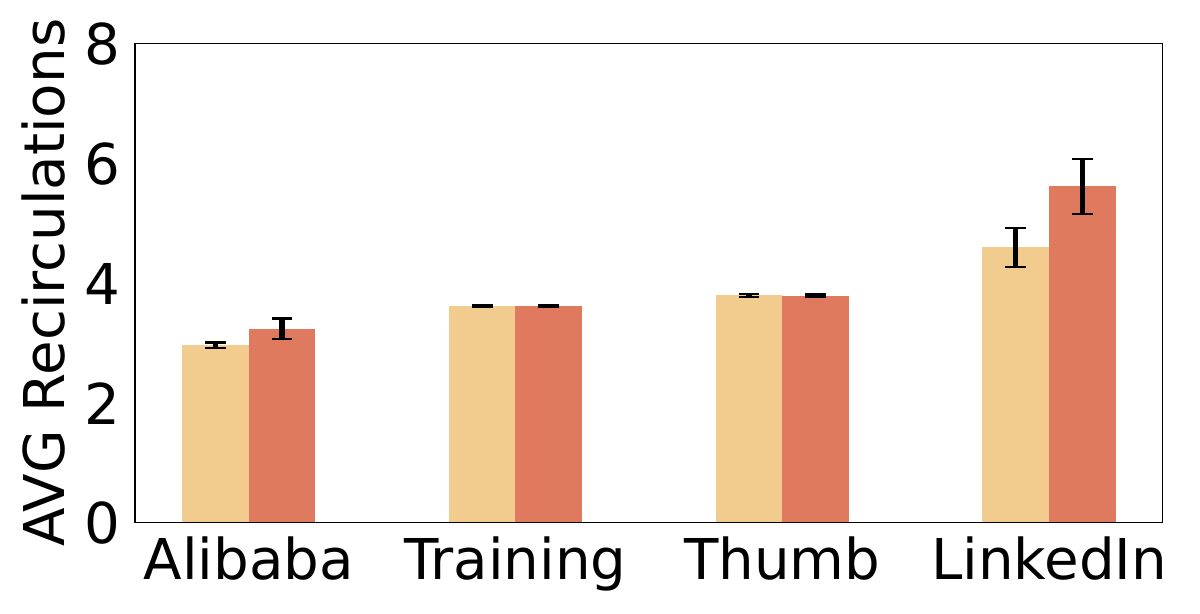}   &
\includegraphics[width=0.49\linewidth]{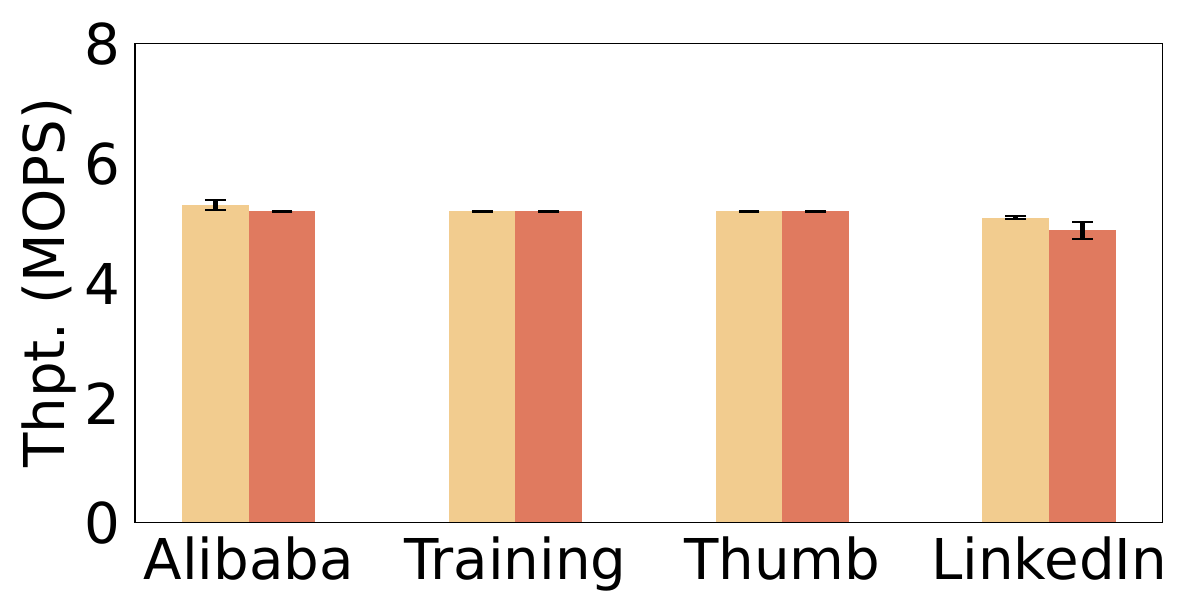}
\vspace{-3pt}\\
\makecell[c]{\small (a) Average recirculation counts} &
\makecell[c]{\small (b) Switch throughput}
\end{tabular}
\vspace{-3pt}
\captionof{figure}{(Exp\#1) Recirculation overhead of Fletch on real workloads.}
\label{fig:exp1overhead}
\vspace{-6pt}
\end{figure}

\noindent{\bf (Exp\#1) Throughput under real-world workloads.}
Figure~\ref{fig:exp1} shows throughput on 16 and 128 simulated servers under four real-world workloads. Under 16 servers, in the standalone mode, \sysname increases
NoCache's throughput by 22.9\%, 51.7\%, 53.7\%, and 47.6\% in Alibaba, Training, Thumb, and LinkedIn, respectively, as in-switch caching absorbs skewed hot-path requests that otherwise cause load imbalance in NoCache. In the complementary mode, \sysnameplus increases CCache's throughput by 8.0\%, 33.5\%, 36.3\%, and 31.5\% in the same workloads, respectively, as CCache forwards all attribute-retrieval requests to servers and cannot resolve server-side load imbalance, while \sysnameplus adds an in-switch caching layer that absorbs these requests before they reach the bottlenecked server.

\sysname has lower throughput than CCache by around 38.1\% across all workloads, as CCache reduces path resolution overhead by caching directory permission metadata at clients and bypasses HDFS lease management and distributed transaction overhead. This is expected: client-side caching and in-switch load balancing are orthogonal optimizations targeting distinct bottlenecks, which is precisely why the complementary mode combines both.

Under 128 servers, the gains in both modes increase substantially. In the standalone mode, \sysname increases NoCache's throughput by 11.0\%, 134.6\%,
181.6\%, and 71.2\% in Alibaba, Training, Thumb, and LinkedIn, respectively. In the complementary mode, \sysnameplus increases CCache's throughput by 14.7\%, 103.6\%, 139.6\%, and 57.3\% for the same workloads, respectively. The gains are significantly higher than in 16 servers (except in Alibaba) as they improve server scalability via load balancing.  The exception in Alibaba is attributed to its high write ratio: under write-through caching, 
cache consistency maintenance for frequent writes incurs substantial overhead that limits scalability gains.
Across both scales, the results confirm that the two modes target distinct deployment regimes: the standalone mode offers sizable improvements over a no-cache deployment, while the complementary mode offers further gains on top of an existing
client-side caching deployment.

We further analyze the recirculation overhead of \sysname on 16 and 128 simulated servers. We measure the {\em recirculation count} (i.e., the average number of recirculations per request) for each workload using the Barefoot Shell monitoring tool \cite{bfshell}. Following the server rotation methodology (\S\ref{subsec:methodology}), in each iteration, a physical machine acts as one simulated server and processes its assigned share of the workload's 32 million requests; we read the number of packets passing through the switch's recirculation port at the end of each iteration. Summing these counts across all iterations and dividing by 32 million yields the recirculation count.

Figure~\ref{fig:exp1overhead}(a) shows the recirculation count results. Across all workloads, the count ranges from 3.00 to 5.61, and includes two components: (i) one mandatory cross-pipeline recirculation per request since the client-connected pipeline and the server-connected pipeline hosting the lock counter arrays are physically separate in our testbed (\S\ref{subsec:methodology}); and (ii) the remaining recirculations from in-switch path resolution (\S\ref{subsec:pathhashing}) for reads and lock acquisition (\S\ref{subsec:lockflow}) for writes. The counts remain modest across all workloads, showing that \sysname's recirculation overhead is manageable under realistic access patterns.

We next evaluate the switch's processing capacity under the measured recirculation counts to quantify the available switch-side headroom. To isolate switch throughput from backend server performance, we pre-load the 5,000 hottest files and their ancestors into the in-switch cache (\S\ref{subsec:methodology}) and use a DPDK-based client \cite{dpdk} to issue \texttt{stat} requests exclusively on these cached paths, so that all requests are served entirely by the switch on cache hits without any server involvement.

A cache-hit \texttt{stat} request on a path at depth $L$ always incurs exactly $L + 2$ recirculations: $L$ for in-switch path resolution (one recirculation per level, for levels $1$ through $L$; \S\ref{subsec:pathhashing}), one for decrementing the lock counter at the final level after path resolution completes, and one cross-pipeline recirculation in our testbed (\S\ref{subsec:methodology}). To reproduce a target recirculation count $r$ from Figure~\ref{fig:exp1overhead}(a), we randomly select a cached path at depth $\lceil r \rceil - 2$, rounding up $r$ to the nearest integer to ensure a conservative evaluation since path depth must be an integer. Under this setup, we increase the client's sending rate until the switch begins dropping packets internally (i.e., the client receives fewer responses than requests sent while its receive buffer does not overflow) and report the highest sustained throughput before this saturation point.

Figure~\ref{fig:exp1overhead}(b) shows the switch's peak throughput under the recirculation counts of the four workloads. \sysname achieves 5.1--5.3~MOPS and 4.9--5.2~MOPS on 16 and 128 servers. The switch peak throughput decreases as the recirculation count increases, indicating that recirculation is a switch-side overhead of \sysname's design. Nevertheless, within the recirculation regime observed under real-world workloads (3.00--5.61 from Figure~\ref{fig:exp1overhead}(a)), the switch sustains a throughput at least 6.3$\times$ and 2.2$\times$ higher than the aggregate throughput of NoCache and CCache on 128 servers (Figure~\ref{fig:exp1}(b)), respectively. This confirms that, under the recirculation counts induced by the evaluated workloads, \sysname and \sysnameplus achieve substantial throughput gains over their respective baselines.
Note that the switch throughput does not exceed the aggregate throughput of \sysnameplus on 128 servers. This exposes a limitation of server rotation: when a system's aggregate throughput approaches the switch's processing capacity, server rotation may overestimate the actual achievable throughput. In Appendix, we evaluate the recirculation overhead under high write ratios.

\begin{figure}[!t]
\centering
\includegraphics[height=12pt]{figs/legend.pdf}
\begin{tabular}{@{\ }c@{\ }c}
\includegraphics[width=0.26\linewidth]{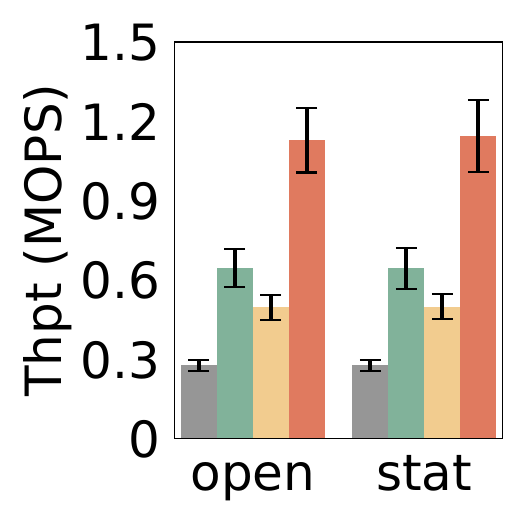}   &
\includegraphics[width=0.675\linewidth]{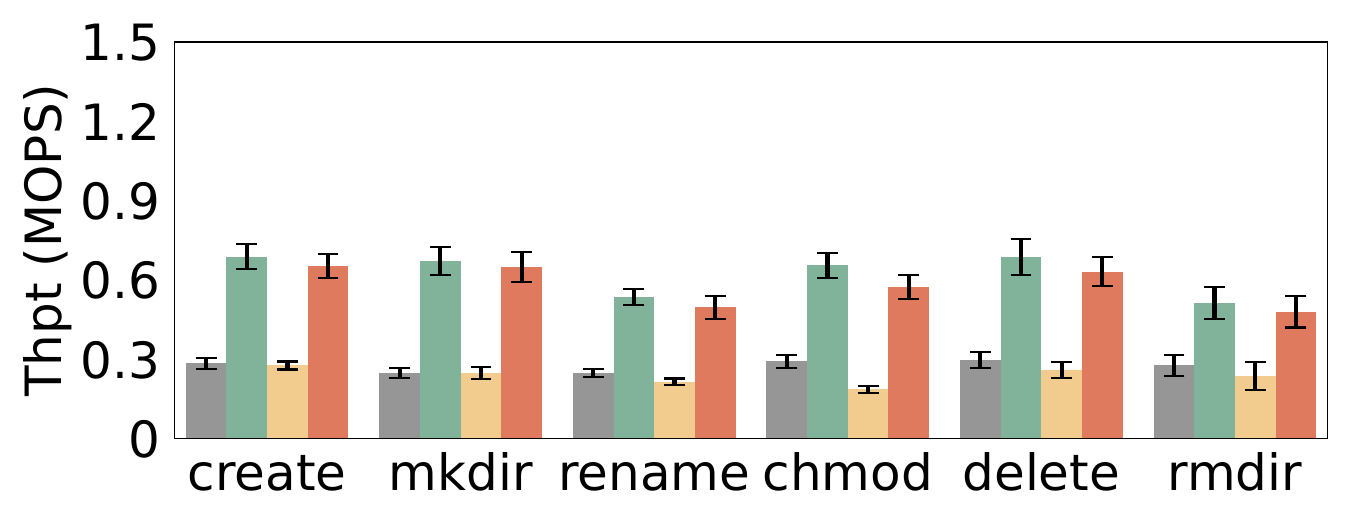}
\vspace{-3pt}\\
\makecell[c]{\small (a) Read operations} &
\makecell[c]{\small (b) Write operations}
\end{tabular}
\vspace{-3pt}
\captionof{figure}{(Exp\#2) Single-operation performance.}
\label{fig:exp2}
\vspace{-6pt}
\end{figure}

\para{(Exp\#2) Single-operation performance.} Figure~\ref{fig:exp2} shows the
throughput of individual metadata operations.
For read operations (\texttt{open} and \texttt{stat}), \sysname increases NoCache's throughput by 80.1\% and 80.5\%, while \sysnameplus increases CCache's throughput by 74.9\% and 77.5\%, respectively.  For write operations (\texttt{create}, \texttt{mkdir}, \texttt{rename}, \texttt{chmod}, \texttt{delete}, and \texttt{rmdir}), \sysname has lower throughput than NoCache by 2.7\%, 0.2\%, 13.2\%, 36.5\%, 12.7\%, and 14.6\%, respectively, while \sysnameplus has lower throughput than CCache by 4.9\%, 3.5\%, 7.2\%, 12.3\%, 8.1\%, and 6.4\%, respectively.  The performance drops stem from the switch's cache maintenance overhead.  Among write operations, \texttt{chmod} incurs the highest overhead as \texttt{chmod} on cached paths requires fetching metadata from HDFS namenodes and updating the in-switch cache; in contrast, \texttt{rename}, \texttt{delete}, and \texttt{rmdir} on cached paths only mark paths as deleted, and \texttt{create} and \texttt{mkdir} only involve creations on new files and directories on HDFS namenodes, respectively, and will not trigger cache updates. Although writes incur overhead due to consistency maintenance, writes are rare in realistic workloads, and both \sysname and \sysnameplus maintain their performance gains (Exp\#1).

\begin{figure}[!t]
\begin{minipage}{0.52\linewidth}
\centering
\includegraphics[width=0.99\linewidth]{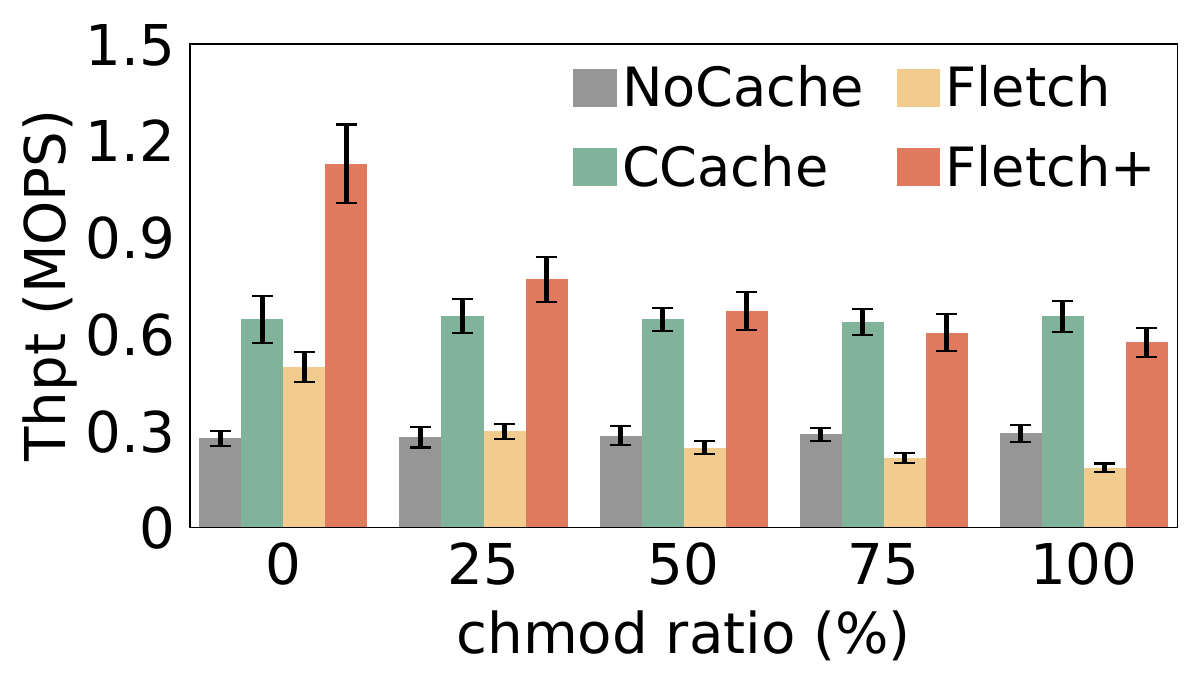}
\captionof{figure}{(Exp\#3) Throughput versus \texttt{chmod} ratio.}
\label{fig:exp3}
\end{minipage}%
\hspace*{0.3em}
\begin{minipage}{0.45\linewidth}
\centering
\Large
\captionof{table}{(Exp\#3) Recirculation count versus \texttt{chmod} ratio.}
\label{tab:exp3}
\resizebox{\linewidth}{!}{
		\renewcommand{\arraystretch}{1.1}
		\begin{tabular}{|c|c|c|c|}
			\toprule[1pt]
			\textbf{\texttt{chmod ratio}}& \textbf{SingleLock} & \textbf{MultiLock} \\
			\bottomrule[0.5pt]
			\toprule[0.5pt]
			0\%     & 7.63        & 7.63  \\
			\midrule[0.5pt]
			25\%    & 442.06     & 37.21 \\
			\midrule[0.5pt]
			50\%   & 436.29      & 39.42 \\
			\midrule[0.5pt]
			75\%    & 369.43      & 25.67 \\
			\midrule[0.5pt]
			100\%       & 1           & 1     \\
			\bottomrule[1pt]
		\end{tabular}
	}
\end{minipage}
\vspace{-6pt}
\end{figure}

\para{(Exp\#3) Impact of \texttt{chmod} ratio.} We further analyze
\texttt{chmod}, which triggers frequent cache updates and shows the most
performance drops in \sysname and \sysnameplus (Exp\#2).  We generate mixed
read-write workloads composed of \texttt{open} (i.e., reads) and
\texttt{chmod} (i.e., writes) with different ratios.  Each of \texttt{open}
and \texttt{chmod} follows a power-law distribution with an exponent of 0.9.
Figure~\ref{fig:exp3} shows that at 0\% \texttt{chmod} ratio, \sysname and
\sysnameplus achieve higher throughput than NoCache and CCache, respectively,
but as the \texttt{chmod} ratio increases, their throughput decreases, while
the throughput of NoCache and CCache remains stable.  \sysname and
\sysnameplus begin to show throughput degradations when the \texttt{chmod}
ratio exceeds 50\%.  At 100\% \texttt{chmod} ratio, \sysname and \sysnameplus
reach the lowest throughput, 36.5\% and 12.3\% lower than NoCache and CCache,
respectively.  Nevertheless, real-world workloads have low \texttt{chmod}
ratios (Table~\ref{tab:workloads}), so \sysname and \sysnameplus still
maintain performance gains in practice (Exp\#1).

\sysname's multi-level read-write locking design (\S\ref{sec:lock}) significantly reduces the overhead in writes.  We compare multi-level locking ({\em MultiLock}) with single-level locking ({\em SingleLock}), which always maps a full path to the first lock counter array.  We measure the recirculation count as described in Exp\#1.  
Table~\ref{tab:exp3} shows that for read-only and write-only workloads, the recirculation counts for both SingleLock and MultiLock are 7.63 and 1, respectively, as \sysname recirculates a read request multiple times for both path resolution and lock counter access, while a write request requires only one recirculation for lock counter access.  MultiLock significantly reduces the recirculation count of SingleLock (e.g., by 93.1\% for 75\% \texttt{chmod}) by mitigating lock contention.

\para{(Exp\#4) Latency analysis.} We analyze request latencies by adjusting
the request sending rate to a target aggregate throughput, as in prior studies
\cite{cheng15,jin17,sheng25,didona19}, so as to analyze the trade-off between
latencies and throughput.  We focus on (i) a read-only workload that issues
32~million \texttt{open} requests and (ii) the Alibaba workload, which has the
largest write ratio (Table~\ref{tab:workloads}). The two workloads show
\sysname's best- and worst-case performance, respectively, under write-through
caching (\S\ref{subsec:goals}).  We follow a power-law distribution with an
exponent of 0.9 and consider 16 simulated servers.

\begin{figure}[!t]
\centering
\includegraphics[height=12pt]{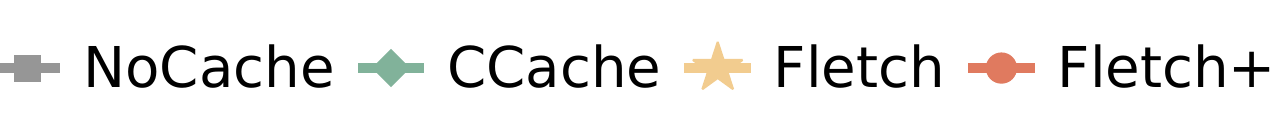}
\vspace{3pt}
\begin{tabular}{@{\ }c@{\ }c}
\includegraphics[width=0.49\linewidth]{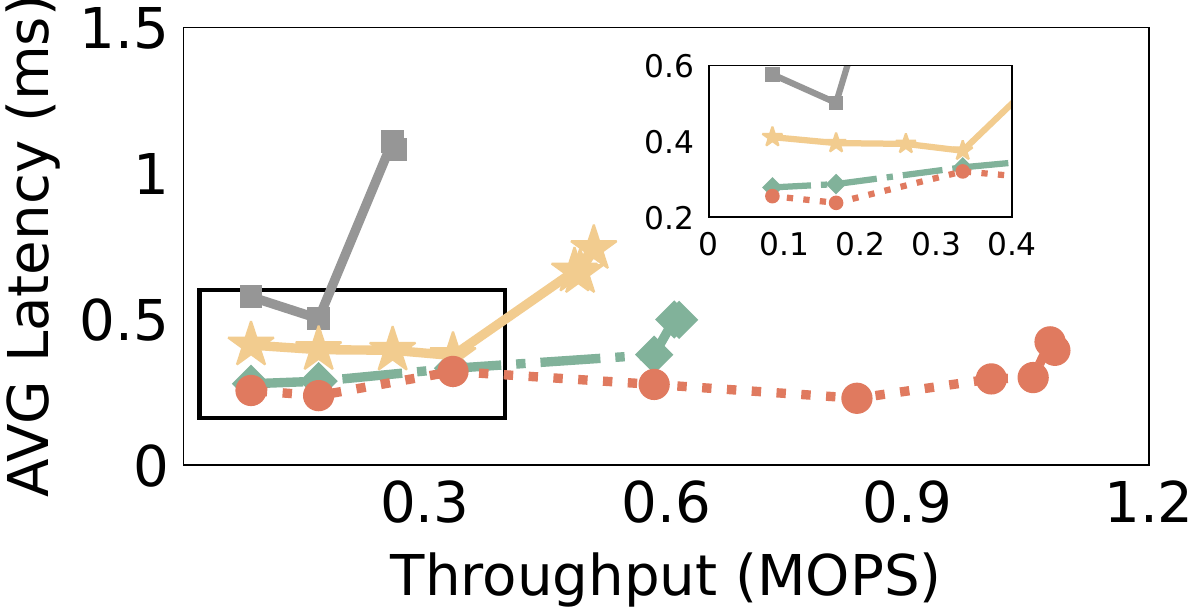} &
\includegraphics[width=0.49\linewidth]{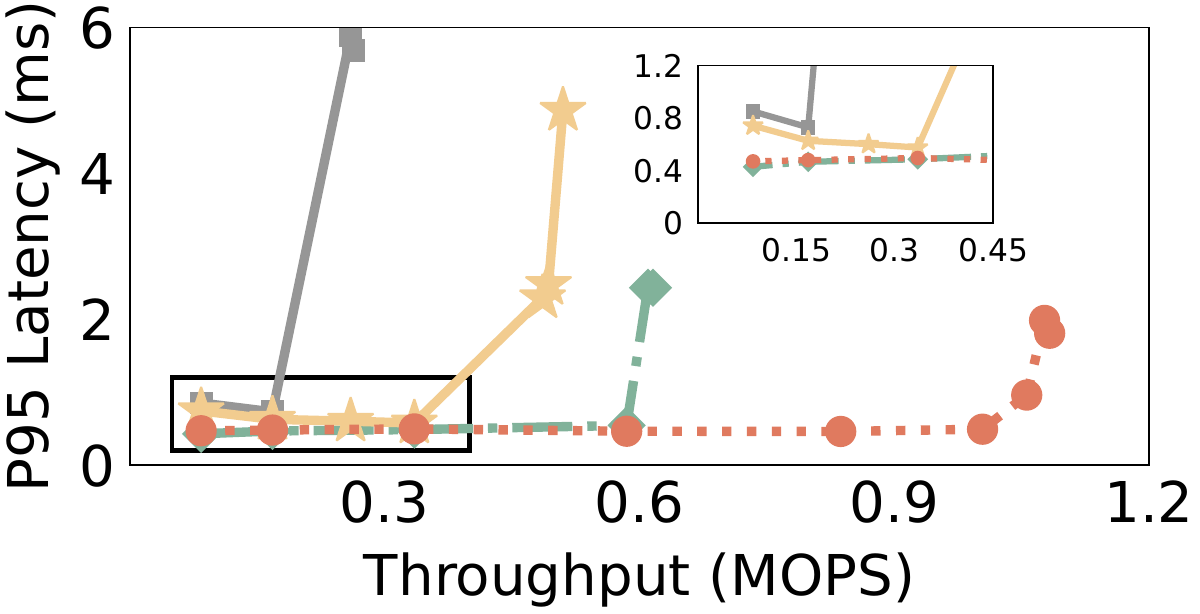}
\vspace{-3pt}\\
\makecell[c]{\small (a) Read-only workload} &
\makecell[c]{\small (b) Read-only workload}
\vspace{3pt}\\
\includegraphics[width=0.49\linewidth]{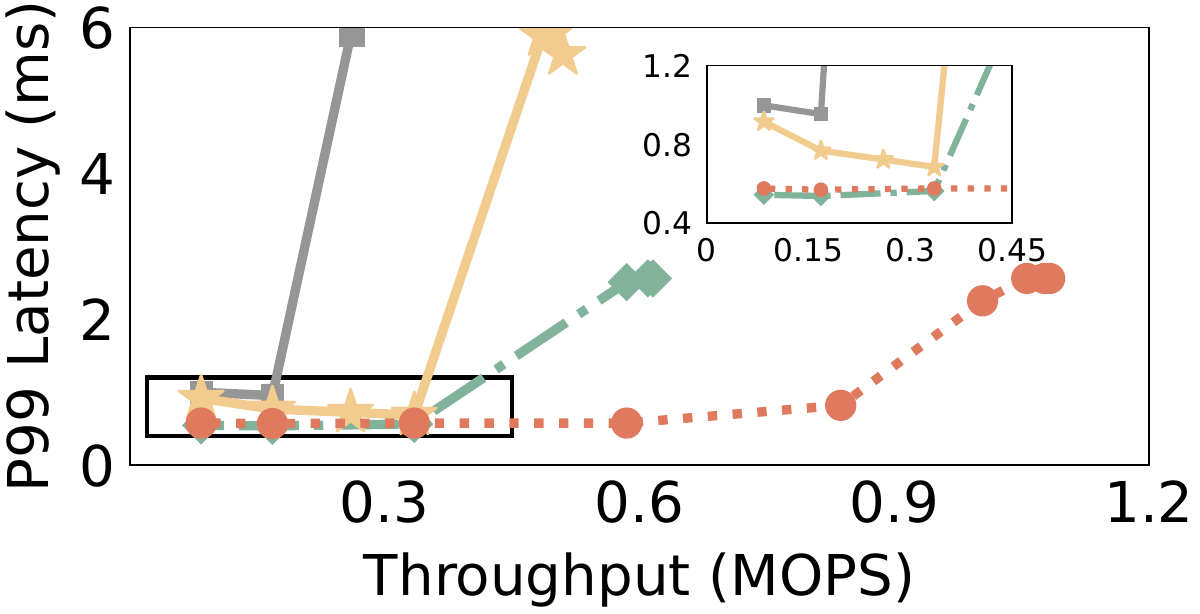} &
\includegraphics[width=0.49\linewidth]{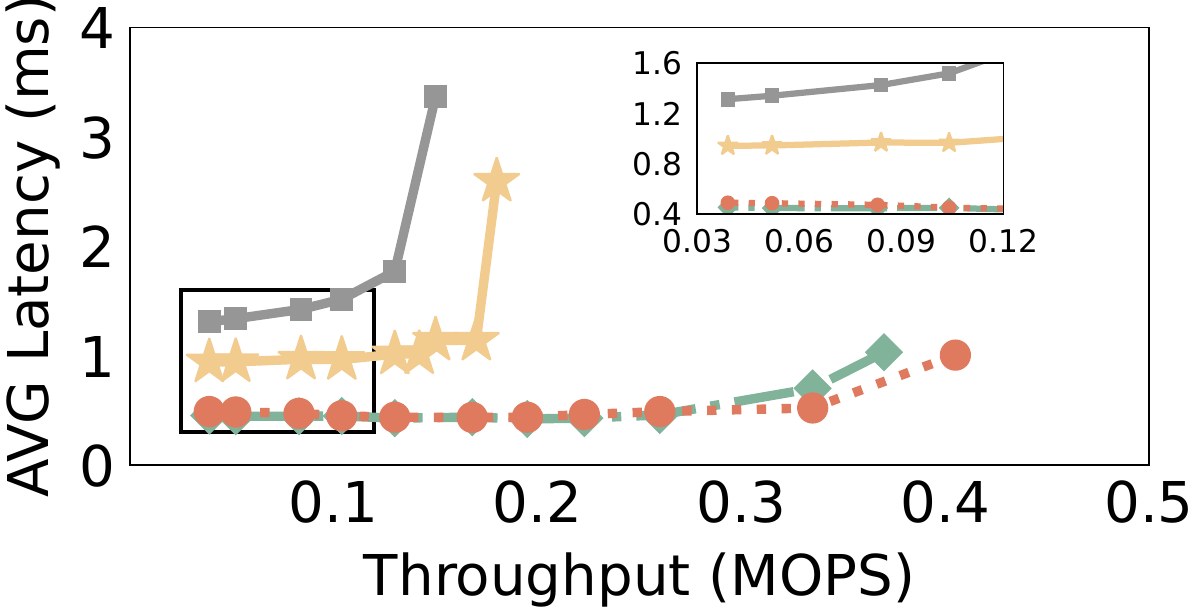}
\vspace{-3pt}\\
\makecell[c]{\small (c) Read-only workload} &
\makecell[c]{\small (d) Alibaba}
\vspace{3pt}\\
\includegraphics[width=0.49\linewidth]{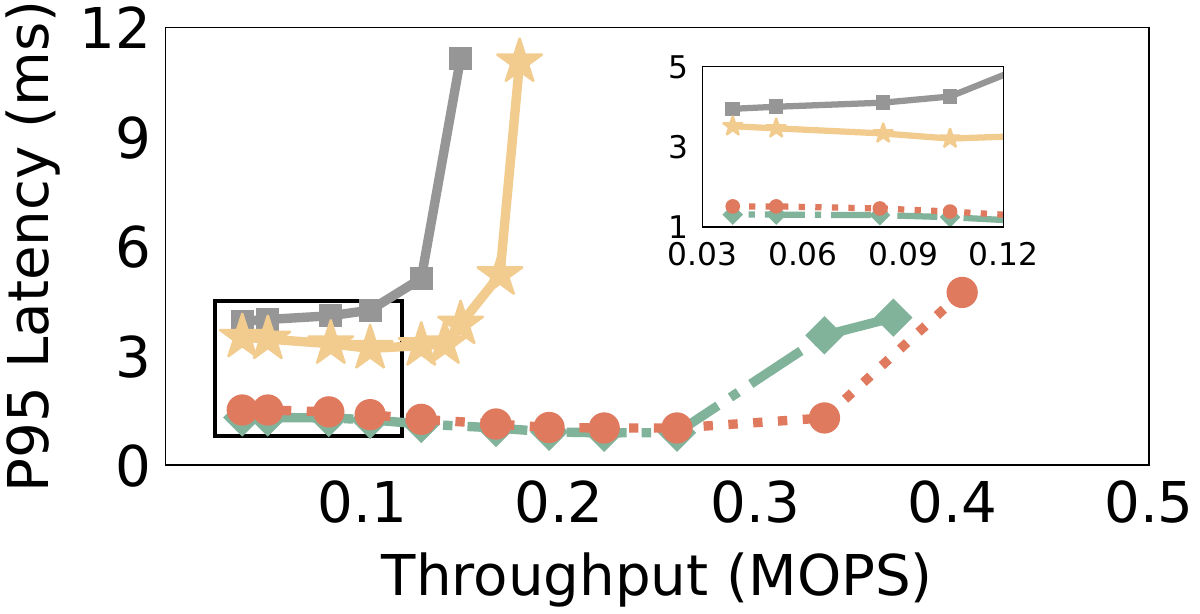} &
\includegraphics[width=0.49\linewidth]{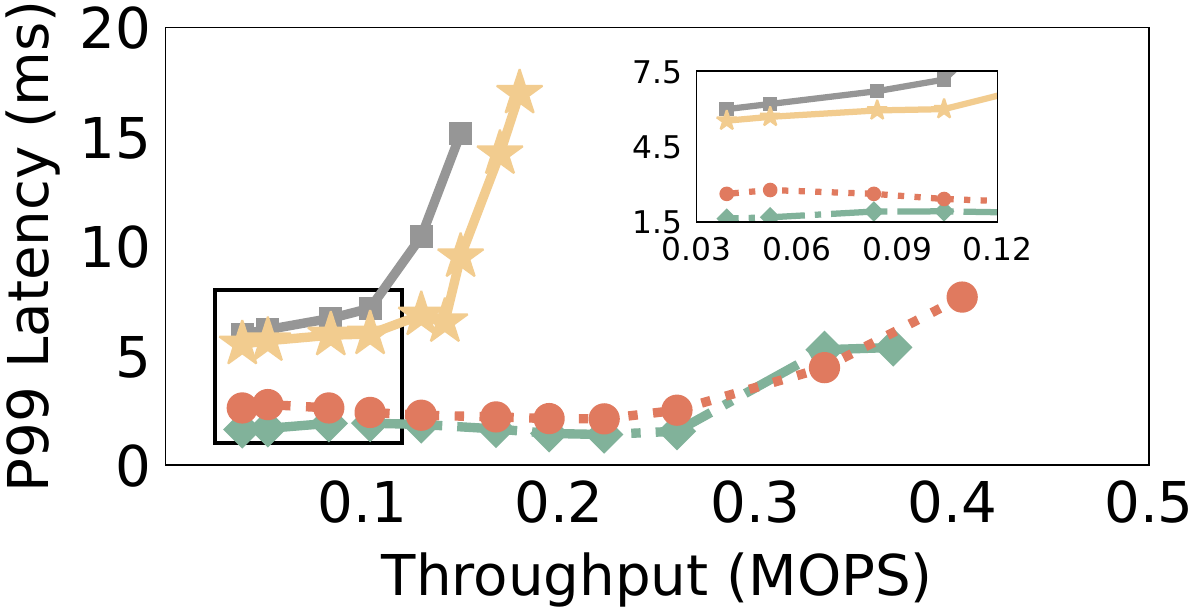}
\vspace{-3pt}\\
\makecell[c]{\small (e) Alibaba} &
\makecell[c]{\small (f) Alibaba}
\end{tabular}
\vspace{-3pt}
\captionof{figure}{(Exp\#4) Latency analysis.}
\label{fig:latency}
\vspace{-6pt}
\end{figure}

Figure~\ref{fig:latency} shows the average, p95, and p99 latency results.
For the read-only workload, all schemes show low latencies at low target
throughput (less than 0.1\,MOPS) as the servers are not saturated and do not
have queueing delays.  When the target throughput increases to 0.2\,MOPS,
NoCache starts to show increasing latencies due to load imbalance under higher
loads.  \sysname maintains low latencies via in-switch caching.  For example,
at target throughput 0.26\,MOPS, \sysname reduces NoCache's average, p95, and
p99 latencies by 64.6\%, 89.8\%, and 87.7\%, respectively.  The same trend is
also observed for \sysnameplus and CCache.  For example, at target throughput 
0.59\,MOPS, \sysnameplus reduces CCache's average, p95, and p99 latencies by
26.9\%, 14.7\%, and 77.0\%, respectively. 

For the Alibaba workload, \sysname also outperforms NoCache. For example, at
target throughput 0.15\,MOPS, \sysname reduces NoCache's average, p95, and p99
latencies by 66.1\%, 65.4\%, and 37.2\%, respectively.  For \sysnameplus and
CCache, at low throughput, they have comparable average and p95 latencies,
while \sysnameplus has slightly higher p99 latencies due to the switch's cache
maintenance overhead.  At high throughput (e.g., 0.35\,MOPS), \sysnameplus
reduces CCache's average, p95, and p99 latencies by 25.5\%, 63.7\%, and
15.5\%, respectively.

\subsection{Impact of Workload Settings}
\label{subsec:microbenchmark}

\begin{figure}[!t]
\centering
\includegraphics[height=12pt]{figs/legend.pdf}
\begin{tabular}{@{\ }c@{\ }c}
\includegraphics[width=0.49\linewidth]{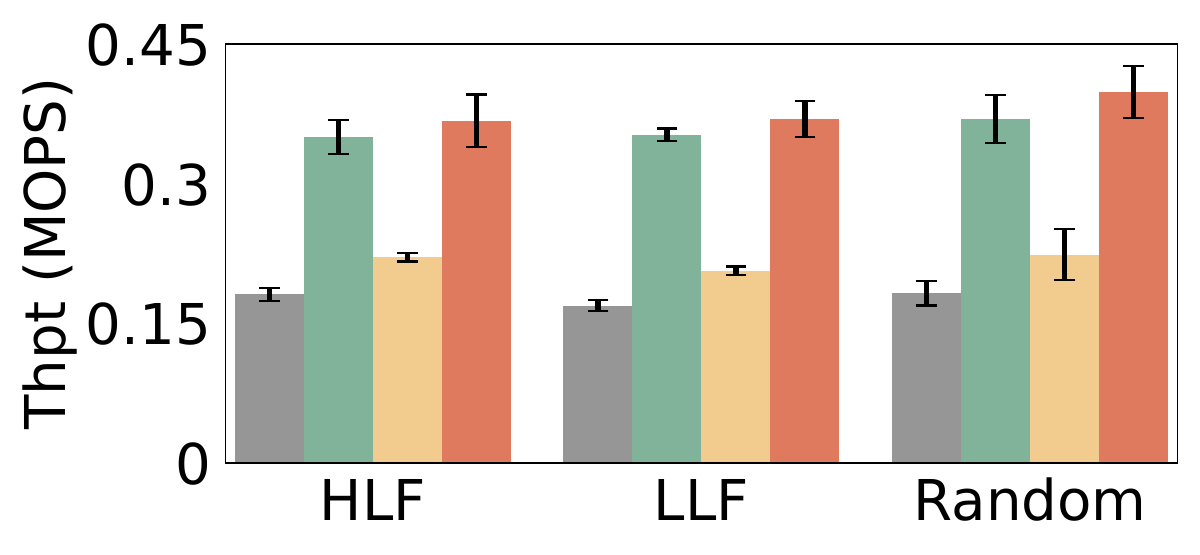}   &
\includegraphics[width=0.49\linewidth]{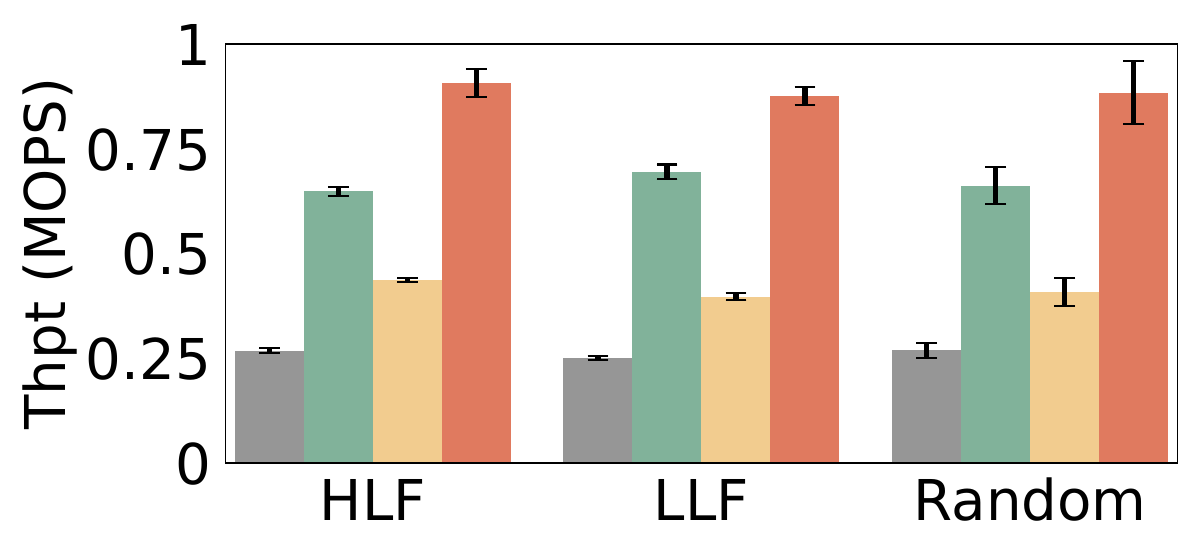}
\vspace{-3pt}\\
\makecell[c]{\small (a) Alibaba} &
\makecell[c]{\small (b) Training}
\vspace{3pt}\\
\includegraphics[width=0.49\linewidth]{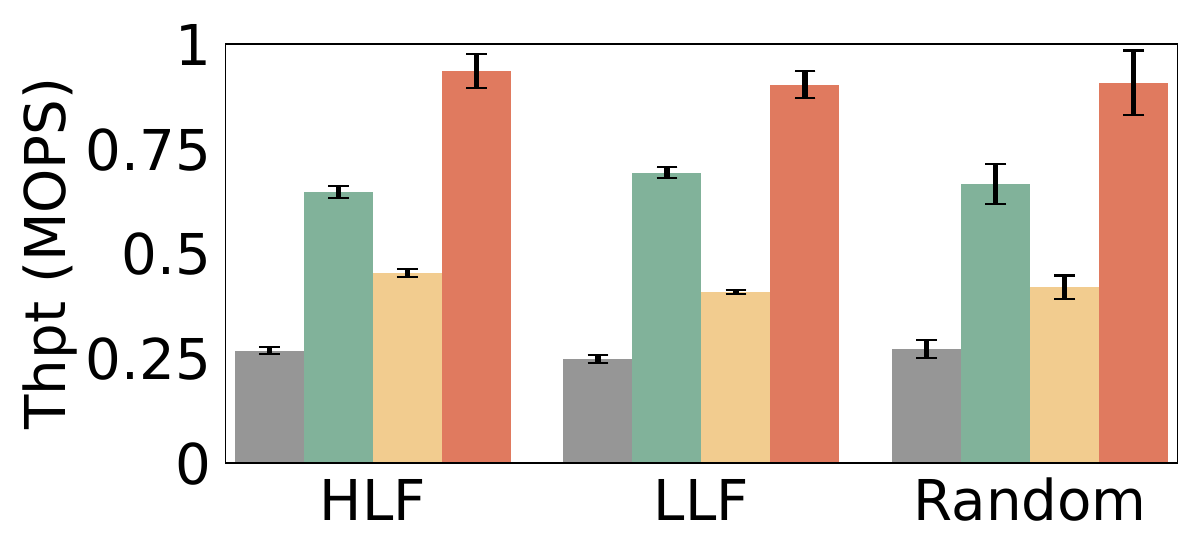}   &
\includegraphics[width=0.49\linewidth]{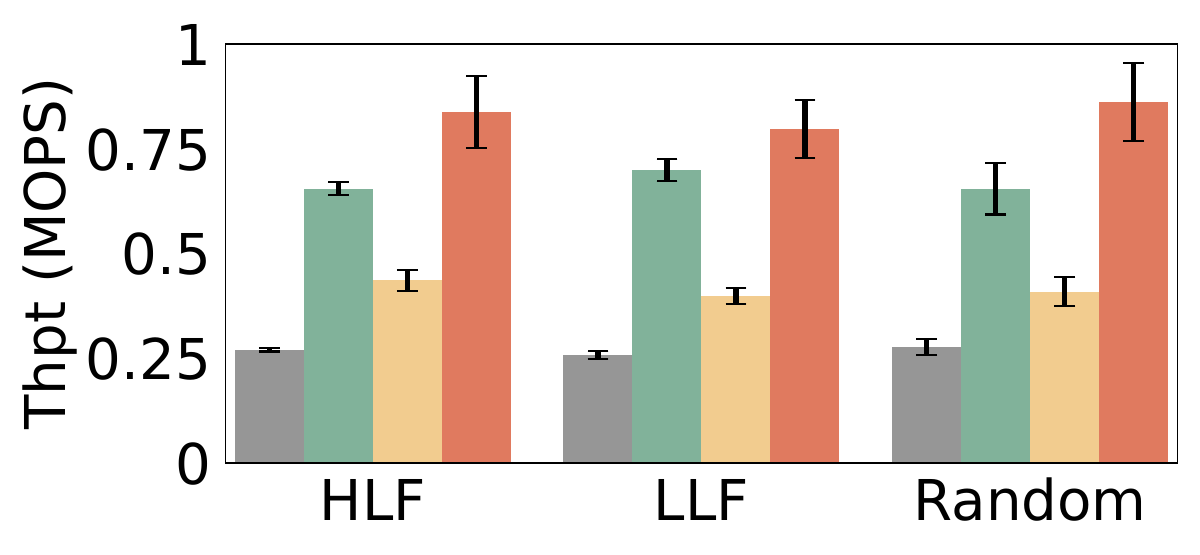}
\vspace{-3pt}\\
\makecell[c]{\small (c) Thumb} &
\makecell[c]{\small (d) LinkedIn}
\end{tabular}
\vspace{-3pt}
\captionof{figure}{(Exp\#5) Impact of file access frequency assignment.}
\label{fig:exp4}
\vspace{-6pt}
\end{figure}

\noindent{\bf (Exp\#5) Impact of file access frequency assignment.}
We generate a sequence of access
frequencies in descending order based on a power-law distribution with an
exponent of 0.9, and another sequence of files based on different sorting
orders.  We assign the $i$-th access frequency to the $i$-th file. We consider
three sorted file sequences: (i) {\em high-level-first (HLF)}, which sorts
files in descending order of their levels (i.e., files at higher levels have
higher access frequencies), (ii) {\em low-level-first (LLF)}, which sorts
files in ascending order of their levels (i.e., files at lower levels have
higher access frequencies), and (iii) {\em random} (our default), which
generates a random sequence of files across all levels.  Figure~\ref{fig:exp4}
shows that \sysname and \sysnameplus still outperform NoCache and CCache,
respectively, under different file access frequency assignments. For example,
in Thumb (the most read-intensive), \sysname and \sysnameplus increase the
throughput of NoCache and CCache by 53.7-69.0\% and 30.2-44.7\%, respectively.

\begin{figure}[!t]
\centering
\includegraphics[height=12pt]{figs/legend.pdf}
\begin{tabular}{@{\ }c@{\ }c}
\includegraphics[width=0.49\linewidth]{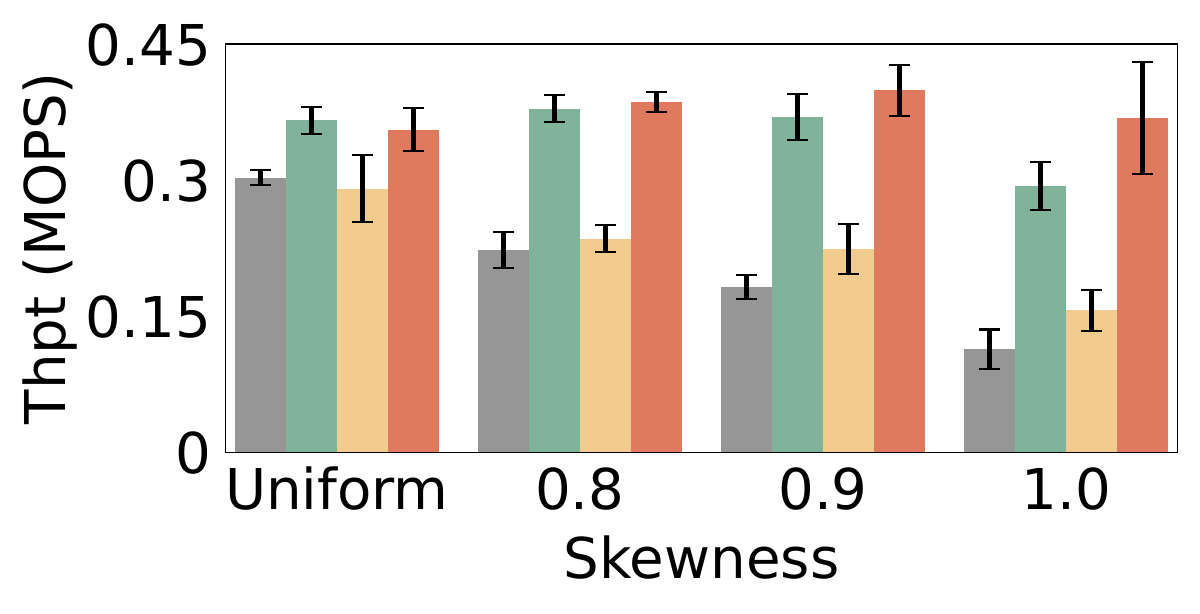}   &
\includegraphics[width=0.49\linewidth]{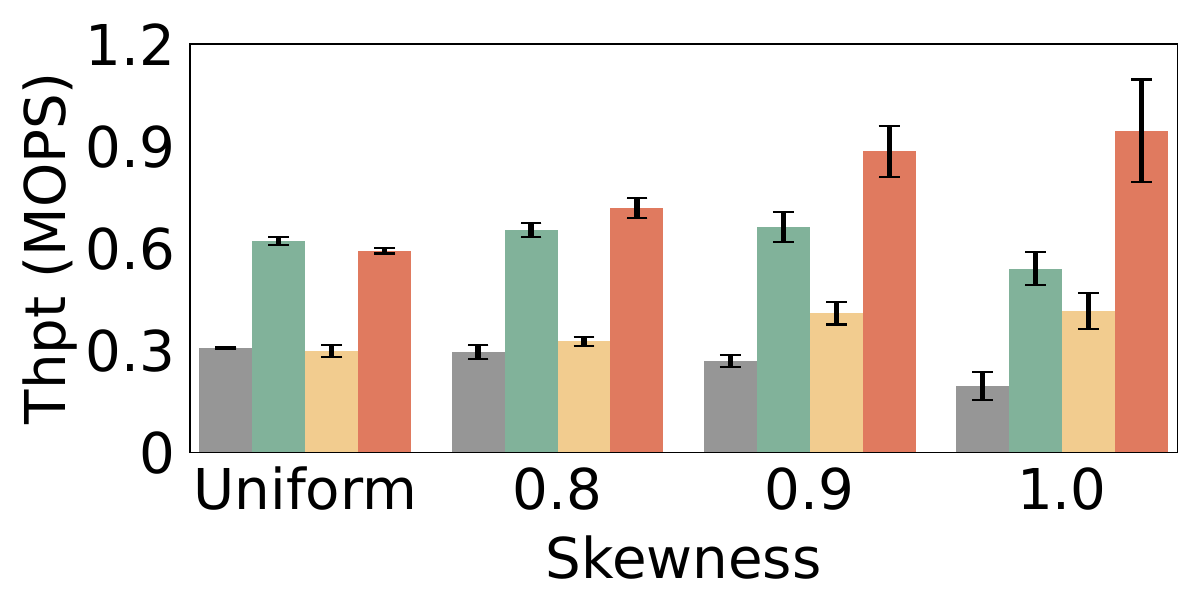}
\vspace{-3pt}\\
\makecell[c]{\small (a) Alibaba} &
\makecell[c]{\small (b) Training}
\vspace{3pt}\\
\includegraphics[width=0.49\linewidth]{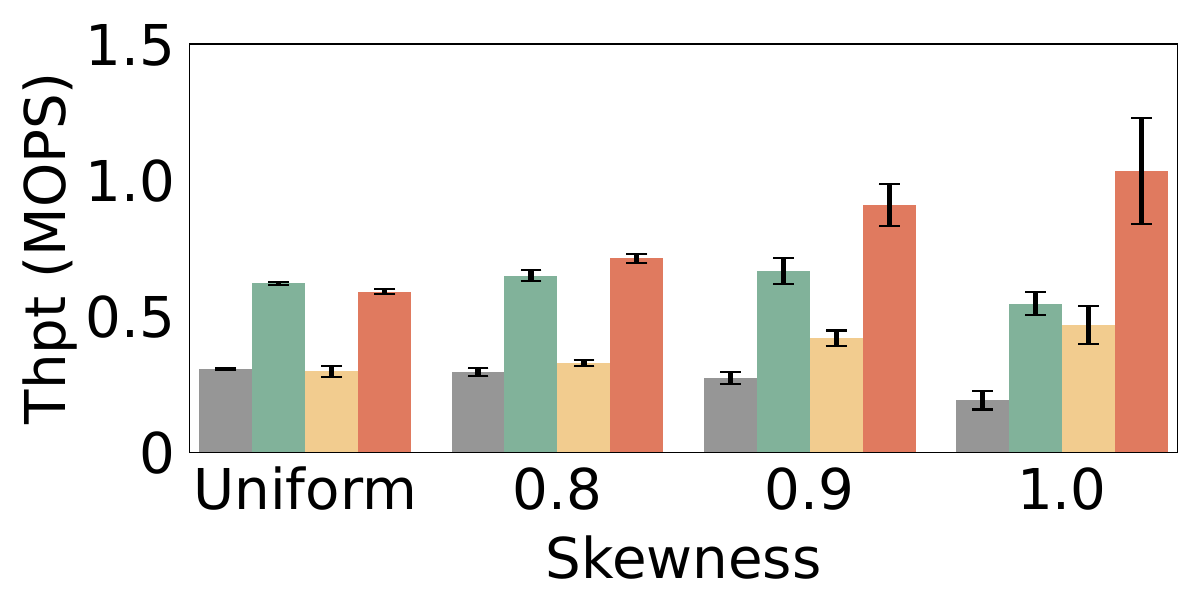}   &
\includegraphics[width=0.49\linewidth]{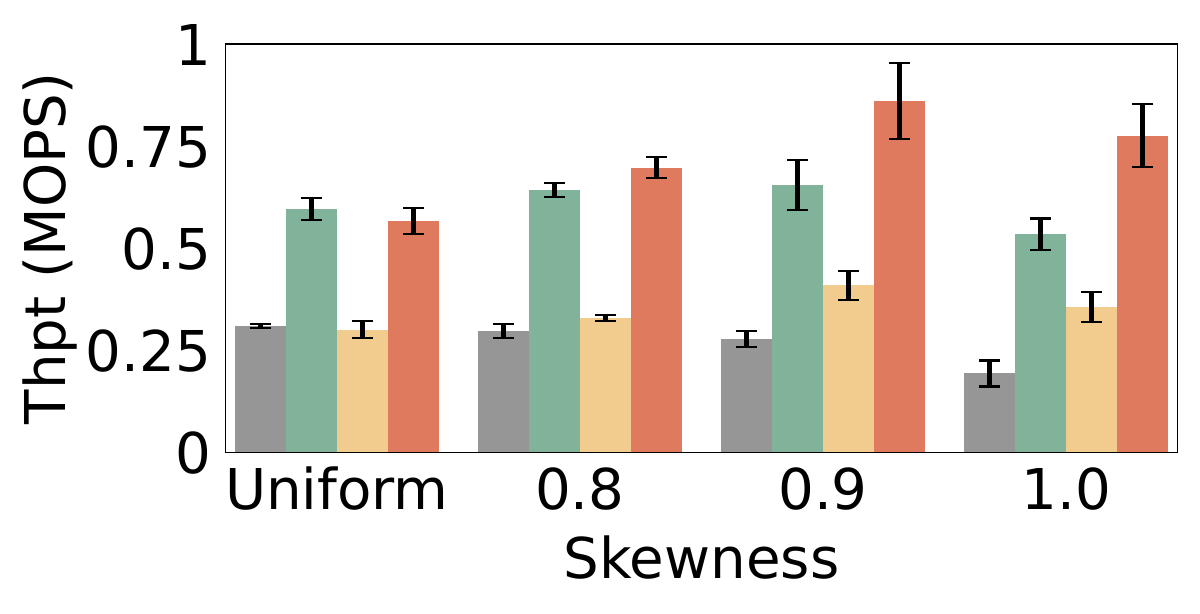}
\vspace{-3pt}\\
\makecell[c]{\small (c) Thumb} &
\makecell[c]{\small (d) LinkedIn}
\end{tabular}
\vspace{-3pt}
\captionof{figure}{(Exp\#6) Impact of access skewness.}
\label{fig:exp5}
\vspace{-6pt}
\end{figure}

\para{(Exp\#6) Impact of access skewness.} We vary the skewness level for
access frequencies under the uniform distribution and varying power-law
distributions with an exponent of 0.8, 0.9 (our default), and 1.0. A larger
exponent implies a more skewed access pattern.  Figure~\ref{fig:exp5} shows
that under the uniform workload, \sysname and \sysnameplus have slightly less
throughput than NoCache and CCache, respectively, by up to 5.0\%, since most
requests come from uncached paths and are served by servers, while \sysname
and \sysnameplus incur extra overhead for maintaining cache consistency. 

For Training and Thumb, for more skewed access, \sysname and \sysnameplus show
increasing throughput via in-switch caching, while NoCache and CCache show
decreasing throughput due to more severe load imbalance.  For example, for
Thumb at exponent 1.0, \sysname and \sysnameplus achieve throughput
gains of 2.4$\times$ and 1.9$\times$ over NoCache and CCache, respectively.

For Alibaba and LinkedIn, as the exponent increases from 0.9 to 1.0, the
throughput of \sysname and \sysnameplus decreases.  Alibaba has the largest
write ratio, while LinkedIn has the highest \texttt{chmod} ratio. They incur
significant maintenance overhead for cache consistency.  Nevertheless, at
exponent 1.0 for LinkedIn, \sysname and \sysnameplus still increase the
throughput of NoCache and CCache by 84.3\% and 45.1\%, respectively.

\begin{figure}[!t]
\centering
\includegraphics[height=12pt]{figs/legend.pdf}
\begin{tabular}{@{\ }c@{\ }c}
\includegraphics[width=0.49\linewidth]{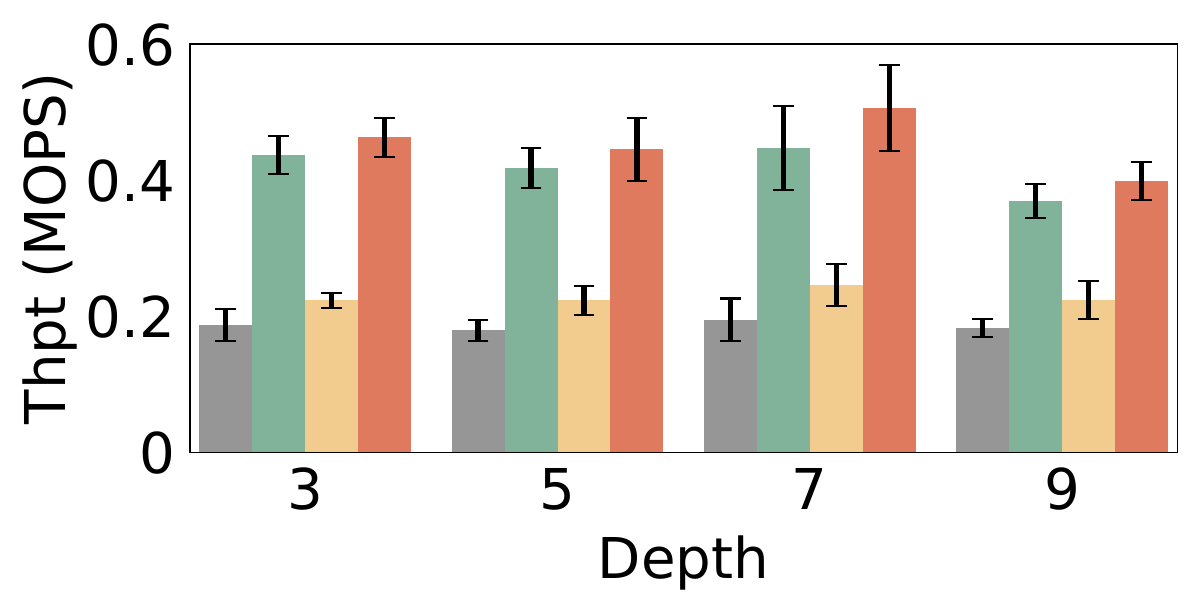}   &
\includegraphics[width=0.49\linewidth]{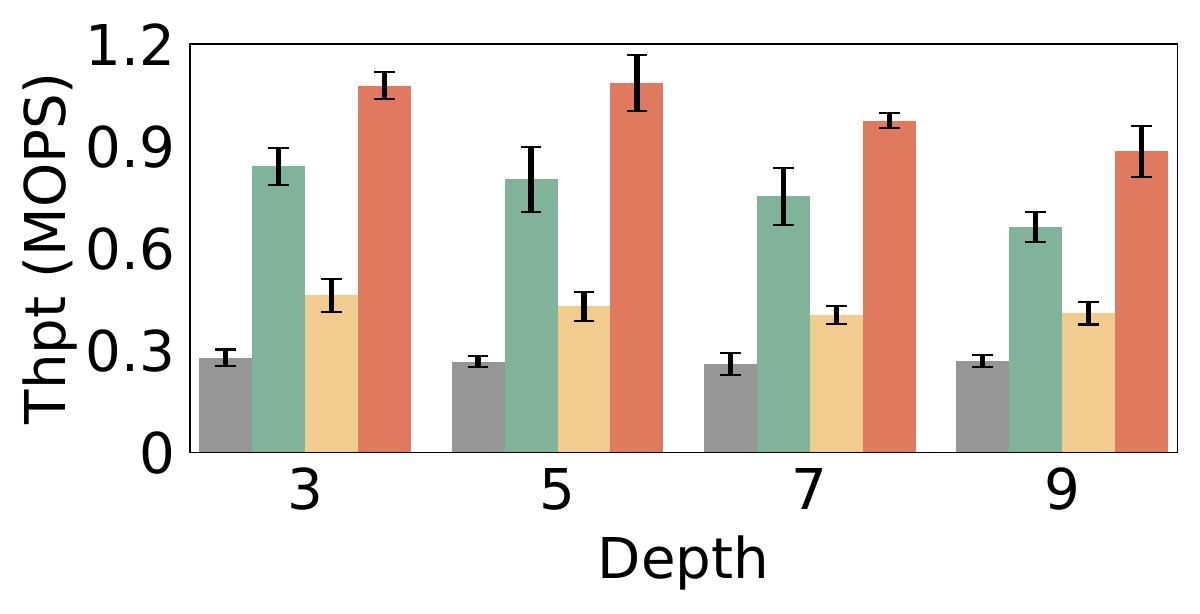}
\vspace{-3pt}\\
\makecell[c]{\small (a) Alibaba} &
\makecell[c]{\small (b) Training}
\vspace{3pt}\\
\includegraphics[width=0.49\linewidth]{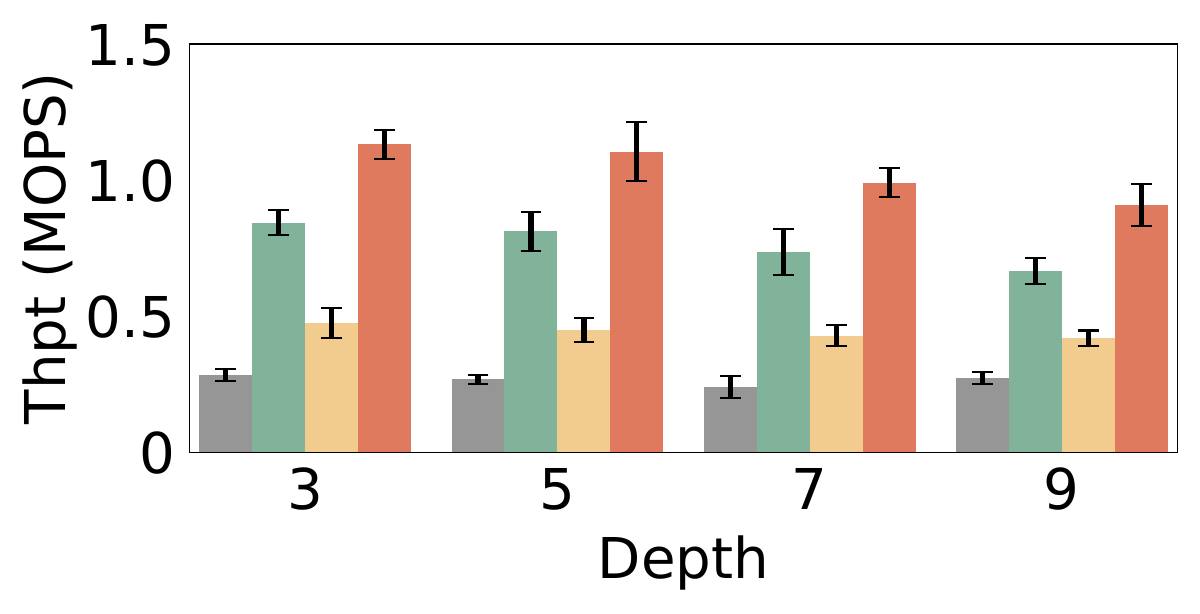}   &
\includegraphics[width=0.49\linewidth]{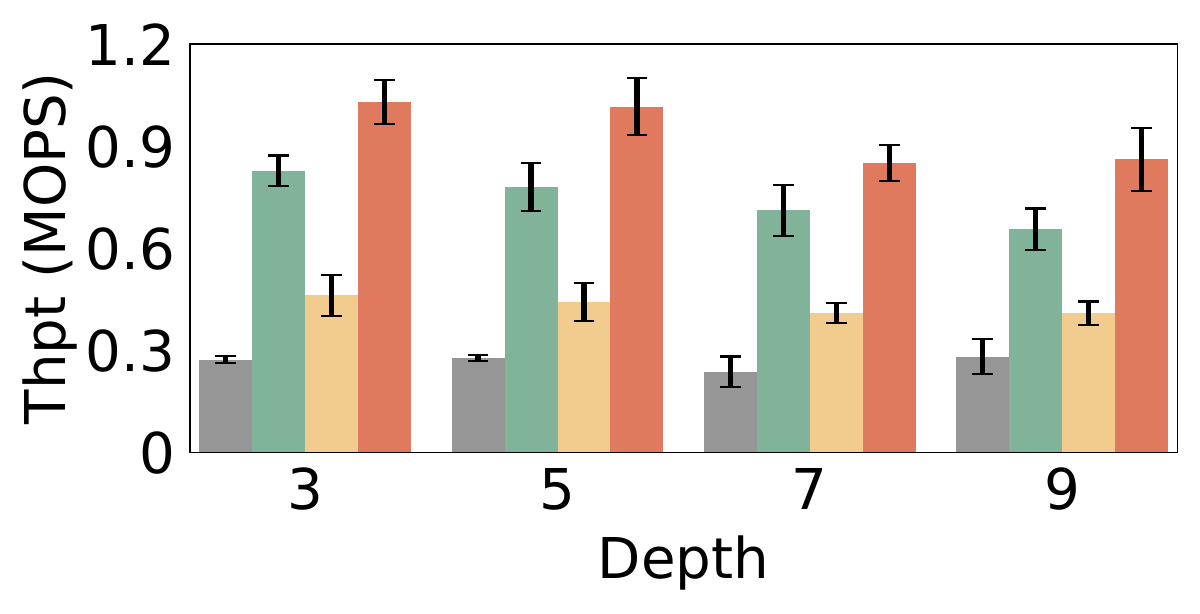}
\vspace{-3pt}\\
\makecell[c]{\small (c) Thumb} &
\makecell[c]{\small (d) LinkedIn}
\end{tabular}
\vspace{-3pt}
\captionof{figure}{(Exp\#7) Impact of maximum path depth.}
\label{fig:exp6}
\vspace{-6pt}
\end{figure}

\para{(Exp\#7) Impact of maximum path depth.}  We vary the maximum path depth
as 3, 5, 7, and 9 (our default).  Figure~\ref{fig:exp6} shows that \sysname
and \sysnameplus always outperform NoCache and CCache, respectively.  For
example, in Thumb, \sysname and \sysnameplus increase the throughput of
NoCache and CCache by 53.7-77.9\% and 34.1-36.3\%, respectively.

\para{(Exp\#8) Impact of dynamic workloads.} We consider dynamic workloads
with varying access frequencies of files over time. We follow the prior
studies \cite{li16, jin17, sheng25} to generate the {\em hot-in} dynamic
pattern, which periodically selects the 100 least-frequently accessed files, 
re-assigns them with the highest access frequencies, and adjusts the access
frequencies of other files accordingly to maintain a power-law distribution.
We set the change period as 20~seconds and run each scheme for 200~seconds to
measure per-second throughput.  We disable server rotation as the system
states change under dynamic workloads; instead, we issue workloads to the
two physical servers and measure performance directly. 

\begin{figure}[!t]
\centering
\includegraphics[height=12pt]{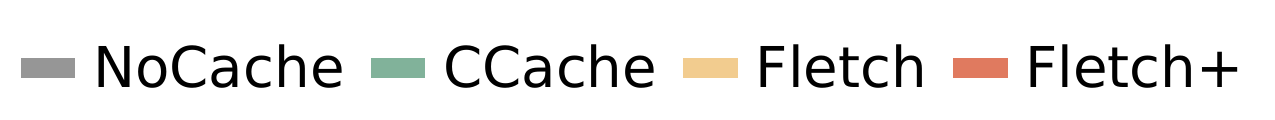}
\begin{tabular}{@{\ }c@{\ }c}
\includegraphics[width=0.49\linewidth]{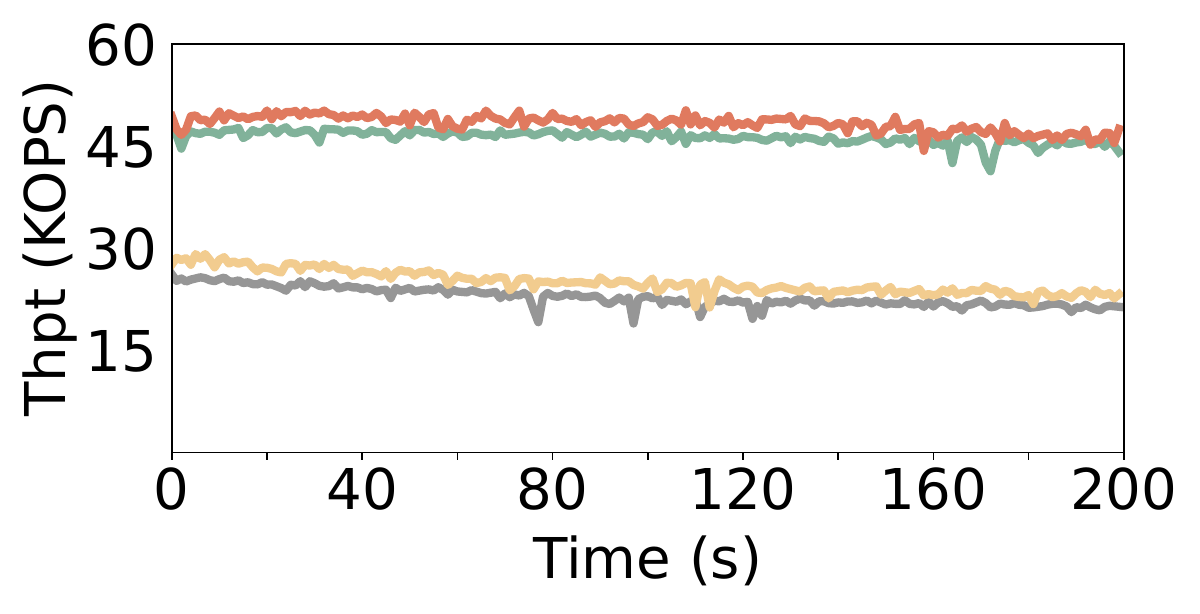}   &
\includegraphics[width=0.49\linewidth]{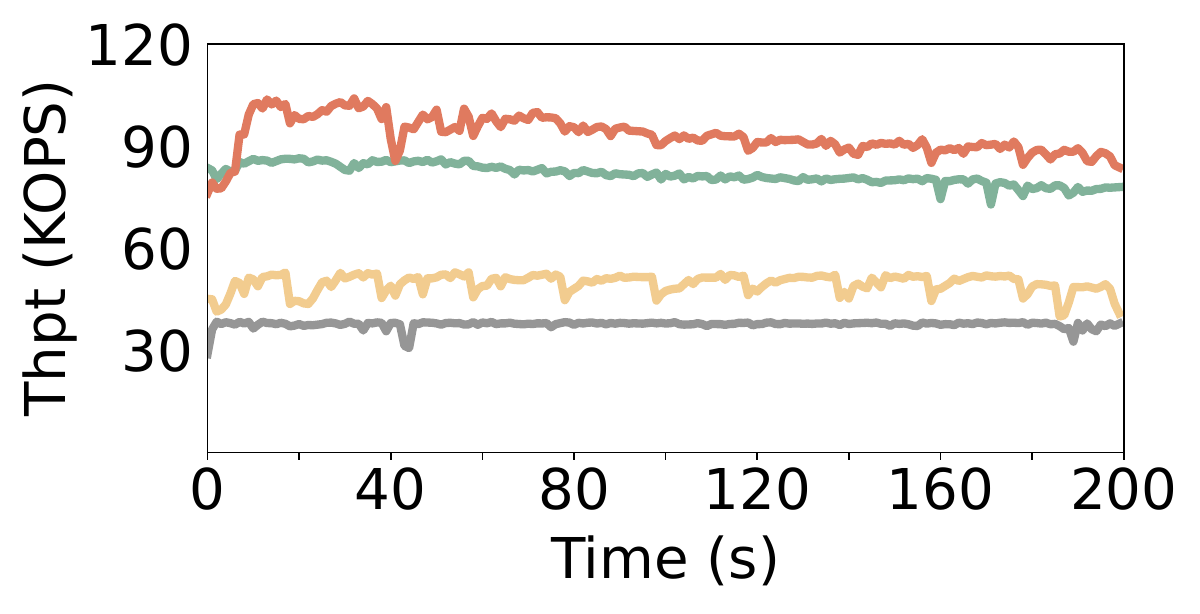}
\vspace{-3pt}\\
\makecell[c]{\small (a) Alibaba} &
\makecell[c]{\small (b) Training}
\vspace{3pt}\\
\includegraphics[width=0.49\linewidth]{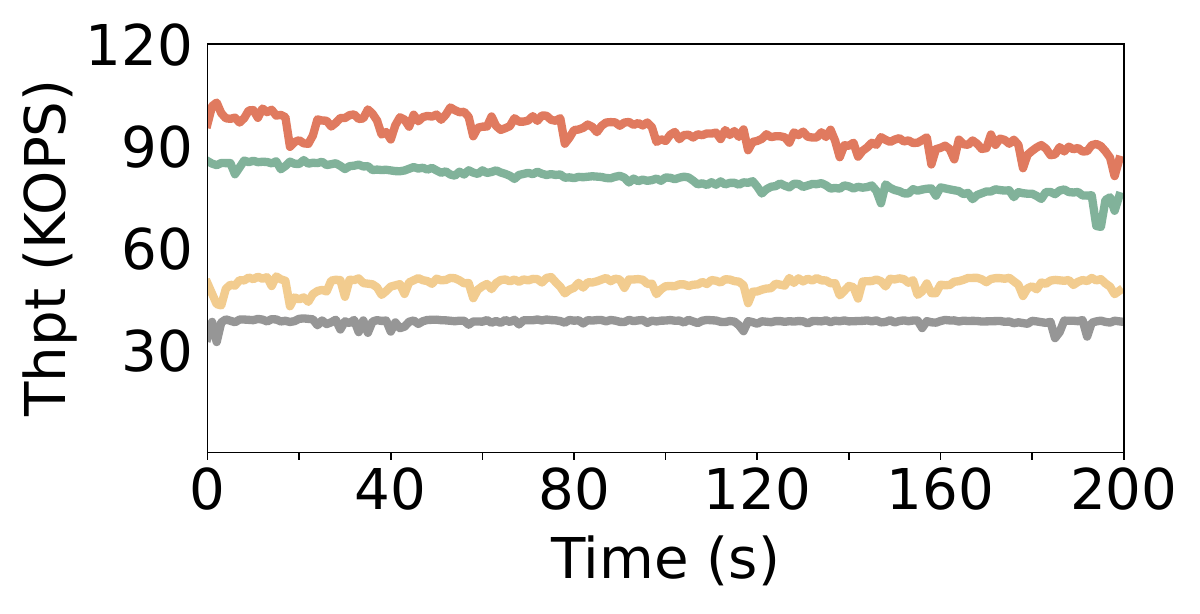}   &
\includegraphics[width=0.49\linewidth]{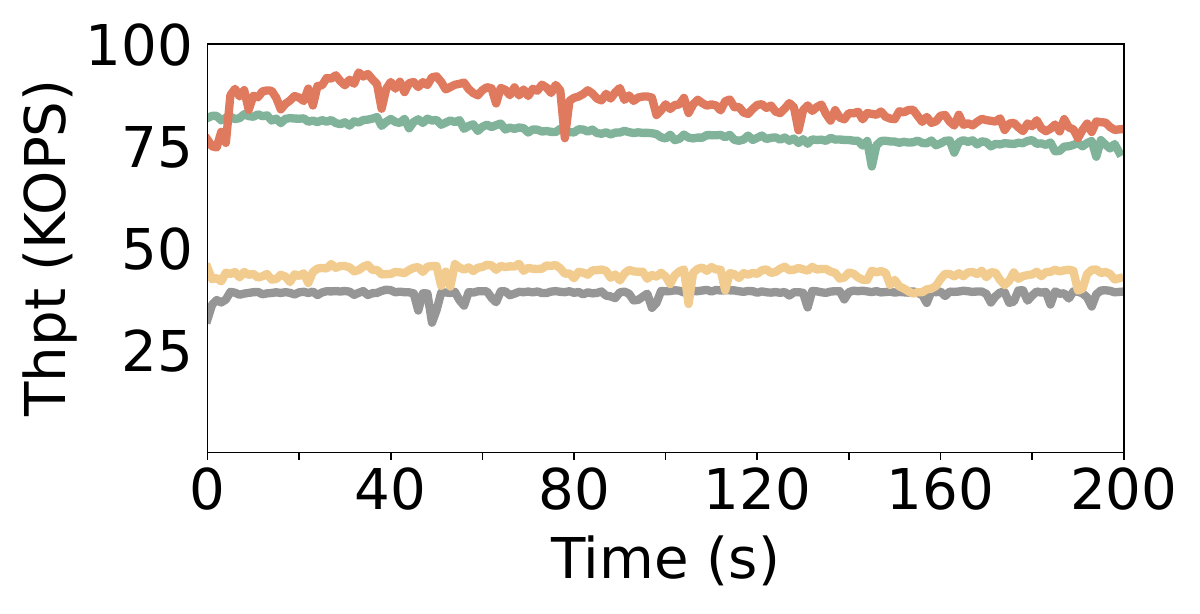}
\vspace{-3pt}\\
\makecell[c]{\small (c) Thumb} &
\makecell[c]{\small (d) LinkedIn}
\end{tabular}
\vspace{-3pt}
\captionof{figure}{(Exp\#8) Impact of dynamic workloads.}
\label{fig:exp7}
\vspace{-6pt}
\end{figure}

Figure~\ref{fig:exp7} shows that for Training, Thumb, and LinkedIn, \sysname
and \sysnameplus show performance dips due to periodic changes of file access
frequencies.  Before new hot records are admitted, performance dips occur,
but \sysname and \sysnameplus quickly admit new hot records and return to high
performance with path-aware cache management
(\S\ref{sec:pathaware}).  Also, local hash collision resolution
(\S\ref{sec:token}) incurs minimal overhead to cache admission and eviction.
For Alibaba, which has the largest write ratio, \sysname and \sysnameplus have
marginal performance gains over NoCache and CCache, respectively,
as they incur cache consistency overhead.
Nevertheless, \sysname and \sysnameplus still effectively respond to dynamic
workloads.

\subsection{Switch Deployment}
\label{subsec:switchusage}

\noindent{\bf (Exp\#9) Switch resource usage.} We measure the resource usage
on the Tofino switch for (i) SRAM (15\,MiB), (ii) number of stages (12), (iii)
number of ALUs (48), and (iv) PHV size (768~bytes). We also quote the resource
usage of two in-switch key-value cache systems, NetCache \cite{jin17} and
FarReach \cite{sheng25}, based on the switch resource usage reported in
the paper \cite{sheng25}.
Table~\ref{tab:exp8} shows that NoCache and CCache use the least resources for
L2/L3 forwarding, while \sysname and \sysnameplus also support L2/L3
forwarding to process packets unrelated to metadata operations at line rate
(3.2\,Tbps for the Tofino switch \cite{switchthpt}).  Additionally, \sysname
and \sysnameplus consume additional resources for caching (e.g., using SRAM
and ALUs to cache metadata, track access frequencies, and maintain lock
counters, and using PHVs to parse metadata requests), yet their resource usage
is comparable to NetCache and FarReach (state-of-the-art in-switch key-value
cache approaches).

\sysname's resource footprint is configurable through lock counter array sharing across path levels. In our current prototype, we already apply this by having path levels 8 and beyond share a single lock counter array, keyed by the level-8 hash component, to reduce stage usage (\S\ref{subsec:lockcounters}). This principle generalizes to shallower levels: adjacent levels can be merged under a single shared array. For example, levels 6, 7, and 8 can share level 6's lock counter array as follows. For a read request at depth 8, the switch increments the counter indexed by the path's level-6 hash key once; during path resolution, it does not decrement this counter when processing levels 6 or 7, and decrements it only once after resolving level 8. For a write request at depth 8 awaiting a lock, the switch checks the same counter indexed by the level-6 hash key. Under this scheme, levels 7 and 8 no longer require dedicated arrays, freeing up pipeline stages and ALUs for other on-switch applications. The sharing configuration is specified at compile time. The trade-off is that the shared counter aggregates active read counts across all covered levels, thereby increasing lock acquisition latency relative to dedicated per-level arrays. In exchange, \sysname retains its full cache depth and hit ratio even with fewer pipeline stages and ALUs, providing a tunable trade-off between write lock acquisition latency and hardware resource efficiency.

\begin{table}[!t]
\captionof{table}{(Exp\#9) Switch resource usage (numbers in brackets refer to
the fractions of used resources over total available ones). Numbers of
NetCache and FarReach are quoted from \cite{sheng25}.}
\label{tab:exp8}
\vspace{-3pt}
\resizebox{\columnwidth}{!}{
\renewcommand{\arraystretch}{1.1}
\begin{tabular}{|c|c|c|c|c|c|}
\toprule[1pt]
Scheme & SRAM (KiB)    & \# Stages  & \# ALUs & PHV size (bytes) \\
\bottomrule[0.3pt]
\toprule[0.3pt]
NoCache  & 288 (1.9\%)   & 4 (33.3\%)        & 0 (0\%)      & 256 (33.3\%)          \\
\midrule[0.5pt]
CCache  & 288 (1.9\%)   & 4 (33.3\%)         & 0 (0\%)     & 256 (33.3\%)        \\
\midrule[0.5pt]
NetCache \cite{jin17} & 7856 (51.2\%)   & 12 (100\%) & 45 (93.8\%) & 528 (68.8\%) \\
\midrule[0.5pt]
FarReach \cite{sheng25} & 8080 (52.6\%) & 12 (100\%) & 45 (93.8\%) & 499 (65.0\%) \\
\midrule[0.5pt]
\sysname   & 8976 (58.4\%) & 12 (100\%) & 47 (97.6\%)     & 712 (92.7\%) \\
\midrule[0.5pt]
\sysnameplus   & 8976 (58.4\%) & 12 (100\%) & 47 (97.6\%)     & 712 (92.7\%) \\
\bottomrule[1pt]
\end{tabular}
}
\vspace{-6pt}
\end{table}

\subsection{System Recovery}
\label{subsec:exp_recovery}

In Appendix, we evaluate crash recovery time under switch, controller, and server failures. We find that switch recovery takes 14.6--75.5\,s for 5\,K--20\,K paths due to sequential cache-admission replay, while controller and server recovery are substantially faster (9.9--38.8\,ms and 0.5--2.1\,s, respectively).

\section{Related Work}
\label{sec:related}

\noindent
{\bf Scaling file-system metadata management.} Metadata partitioning
distributes file-system management across multiple servers for scalability. 
There are two primary approaches: (i) {\em dynamic sub-tree partitioning},
which distributes namespace sub-trees across servers 
\cite{weil06,sevilla15,wang21}, and (ii) {\em hash-based partitioning}, which 
distributes file-system metadata across servers via hashing
\cite{patil11,thomson15,niazi17,pan21,liao23,wang23}.  \sysname complements
these approaches as an in-switch cache that absorbs operations upstream of
file-system metadata layers.

\para{Client-side caching.} PanFS \cite{welch08} caches file and directory
metadata and provides callbacks for cache consistency.  IndexFS \cite{ren14}
and LocoFS \cite{li17} use lease-based caching for metadata management and
invalidate cache entries upon lease expiration, but incur high overhead for
renewing cache entries' leases.  InfiniFS \cite{lv22} applies lazy
invalidation for directory access metadata to limit the overhead of
lease-based caching.  Client-side caching often incurs high client-side
complexity and overhead in maintaining cache consistency across a large number
of clients.  \sysname simplifies cache consistency by caching file-system
metadata in a programmable switch that lies on the critical paths of
multiple clients.

\para{In-switch caching.} Programmable switches have been extensively studied for concurrency control \cite{li17eris,zhu19}, network monitoring \cite{gupta18}, replication coordination \cite{jin18}, remote procedure calls \cite{zhao23}, key-value stream aggregation \cite{he23}, and distributed lock management \cite{yu20,zhang24}.  Several studies explore in-switch write-through caching: SwitchKV \cite{li16} and NetCache \cite{jin17} target read-intensive workloads; DistCache \cite{liu19} supports caching across multiple switches; OrbitCache \cite{kim25} supports variable-length items.  Pegasus \cite{li20} and TurboKV \cite{eldakiky20} cache replica-to-server mappings for replica selection.  Mind \cite{lee21} caches object-to-memory mappings for disaggregated memory systems.  Concordia \cite{wang21concordia} tracks cache copy locations and states to address concurrency in shared memory systems.  ICH \cite{chen25} and HarmonyCache \cite{chen26} design write-back caching for dynamic workloads and multi-switch read-write separation, respectively. FarReach \cite{sheng25} designs fault-tolerant write-back caching for write-intensive workloads, and its extension, DistReach \cite{sheng25}, supports multiple switches.  The above in-switch caches target key-value stores, which differ from file-system semantics. SwitchFS \cite{xu24} proposes in-switch tracking of directory updates, while maintaining a client-side metadata cache for path resolution.  In contrast, \sysname moves file-system metadata caching to switches, and also complements client-side caching.

\section{Conclusion}

\sysname is an in-switch file-system metadata caching framework, aiming to
achieve high throughput and load balancing for distributed file-system
metadata services. It employs path-aware cache management,
multi-level read-write locking, and local hash collision resolution.
Experiments on a Tofino-switch testbed show that
\sysname achieves significant throughput gains and complements client-side
caching. 

\bibliographystyle{IEEEtranS} 
\bibliography{reference_with_url}

\input{supp}

\end{document}

%% file: supp.tex
\appendix

We report two additional experiments: (i) the recirculation overhead
under higher write ratios (Exp\#S1) and (ii) system recovery time (Exp\#S2).

\section*{Additional Experiments}
\label{sec:extraexp}

\begin{figure}[!t]
\centering
\includegraphics[width=0.9\linewidth]{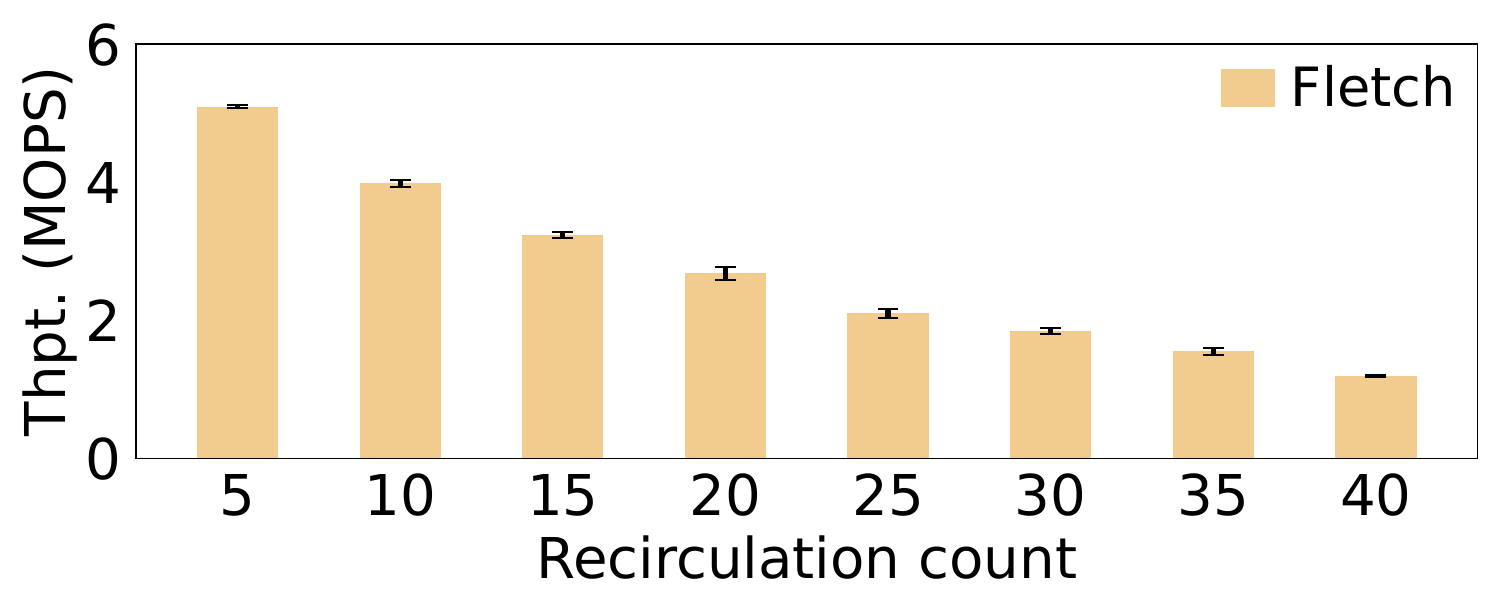}
\captionof{figure}{(Exp\#S1) Switch processing throughput of Fletch under high
recirculation counts.}
\label{fig:exps1}
\end{figure}

\para{(Exp\#S1) Switch processing throughput of \sysname under high
recirculation counts.} We evaluate the switch's processing capacity under
recirculation counts beyond the typical range observed in real-world workloads
(Exp\#1). We focus on \sysname in this experiment. We vary the per-request
recirculation count from 5 to 40. The upper bound of 40 represents the
worst case under high write ratios: Exp\#3 mixes read and write requests
across the full ratio spectrum (0\% to 100\% \texttt{chmod}) and observes
a peak recirculation count of 39.42 at 50\% \texttt{chmod}, beyond which
the recirculation count decreases as fewer concurrent reads contend
with writes for locks.

Following the methodology of switch overhead analysis in Exp\#1,
we pre-load cached paths into the
in-switch cache and use a DPDK-based client \cite{dpdk} to issue
\texttt{stat} requests exclusively on these paths, so that all requests
are served entirely by the switch. As illustrated in Exp\#1, a cache-hit
\texttt{stat} request at depth $L$ incurs $L+2$ recirculations. For a maximum
path depth of 9 (our evaluation default), the natural recirculation count is
at most 11, far below our target of 40. To reach higher recirculation counts,
we extend the \sysname's switch to accept a target recirculation count $r$ as
a per-request input. Before the \texttt{stat} request on a path at depth $L$
enters the normal in-switch caching pipeline, the switch performs
$r - (L+2)$ extra recirculations with no caching logic; after these
extra recirculations, the request follows the normal in-switch caching
processing. We select a cached path at depth 3 across all $r$ values.
For each recirculation count $r$, we increase
the client's sending rate until the switch begins dropping packets
internally and report the highest sustained throughput.

Figure~\ref{fig:exps1} shows the switch throughput across recirculation
counts. The throughput decreases gradually from 5.1\,MOPS at the recirculation
count of 5 to 1.2\,MOPS at the recirculation count of 40, but never collapses.
Even at the recirculation count of 40, the switch throughput remains
4.0$\times$ higher than the aggregate throughput of NoCache on 16 simulated
servers (around 0.3\,MOPS in Exp\#3). Thus, recirculation does not cause
data-plane congestion on 16 simulated servers, even under recirculation counts
(e.g., 40) far beyond those observed in real-world workloads (at most 5.61 in
Exp\#1).

We acknowledge that under write-intensive workloads, the recirculation
overhead becomes significant.  How to mitigate the recirculation overhead in
\sysname is posed as future work. 

\para{\bf (Exp\#S2) System recovery time.} We evaluate the crash recovery time of \sysname and \sysnameplus under switch, controller, and server failures for varying numbers of pre-loaded paths (from 5K to 20K). We first pre-load a given number of paths by triggering cache admission, which populates the in-switch cache, the server's path-token map, and the controller's path-token and hash-token maps (\S\ref{subsec:tokenhash}). We then manually kill a target component to mimic a failure, and trigger the recovery procedure (\S\ref{subsec:recovery}). For controller and server failures, the restarted component passes a recovery flag at startup: the controller reconstructs its maps from its local historical log, while the server queries the controller for its path-token entries in the active log. For a switch failure, we reboot the switch to reset its data-plane state. We run a script to replay cache admission for each path in the active log, so that the controller re-fetches the metadata and re-installs both the metadata and the hash-token map into the switch, with each path retaining its original token.

\begin{figure}[!t]
\centering
\begin{tabular}{@{\ }c@{\ }c@{\ }c}
\includegraphics[width=0.32\linewidth]{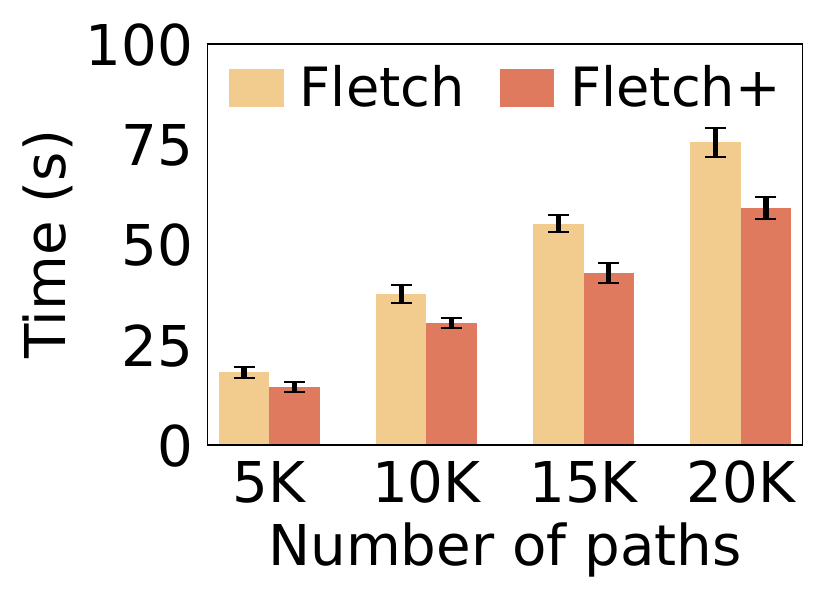}   &
\includegraphics[width=0.32\linewidth]{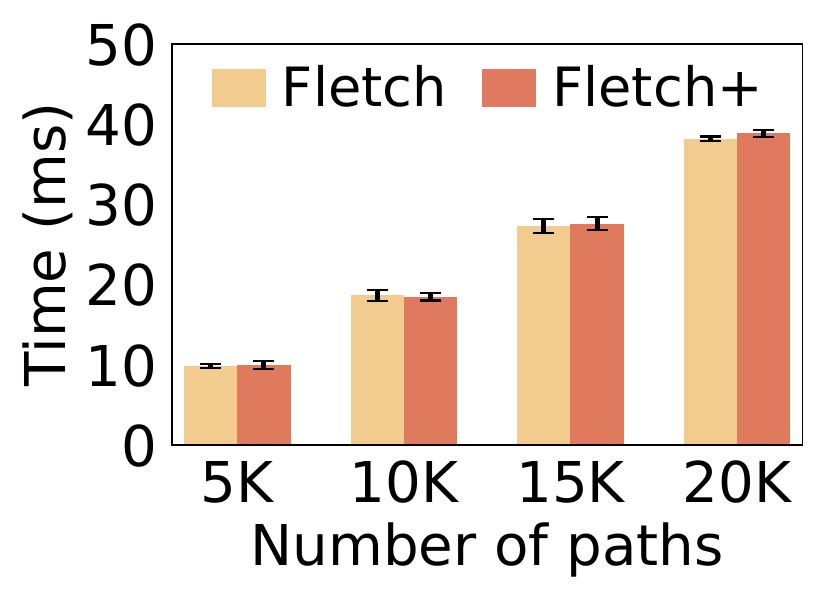}  &
\includegraphics[width=0.32\linewidth]{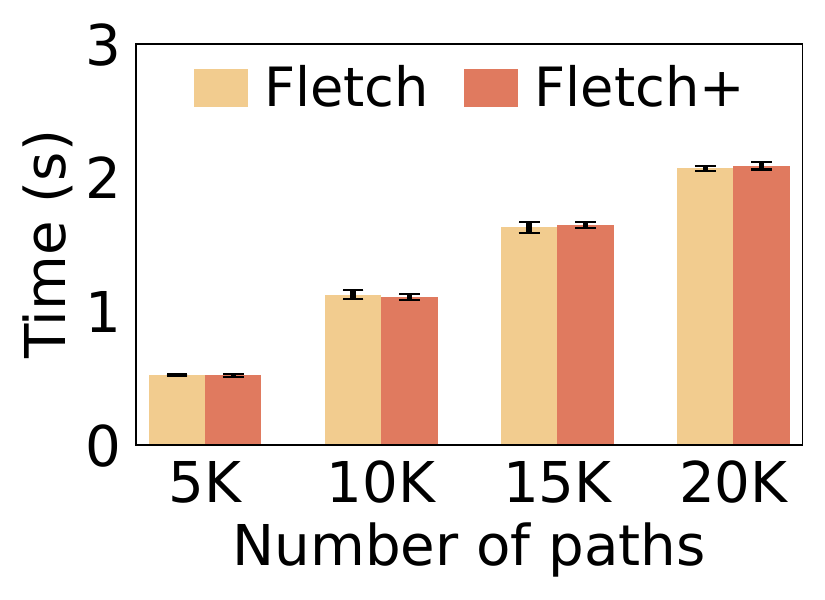}
\vspace{-3pt}\\
\makecell[c]{\small (a) Switch} &
\makecell[c]{\small (b) Controller} &
\makecell[c]{\small (c) Server}
\end{tabular}
\vspace{-3pt}
\captionof{figure}{(Exp\#S2) System recovery time.}
\label{fig:exp10}
\end{figure}

Figure~\ref{fig:exp10} shows the recovery time for each component. The recovery time grows roughly linearly with the number of paths, as each path is processed sequentially. From 5K to 20K paths, server recovery takes 0.5\,s to 2.1\,s, controller recovery takes 9.9\,ms to 38.8\,ms, and switch recovery takes 14.6\,s to 75.5\,s. Switch recovery is the slowest as it replays cache admission for each path, which involves fetching metadata from the backend and installing it along with the hash-token map into the switch. Controller recovery is the fastest as it only reads the local historical log without network communication. Server recovery falls in between, as the controller transmits the active log via UDP, incurring network round-trips.

\sysnameplus recovers 21.6\% faster than \sysname on switch recovery at 20K paths, because switch recovery requires fetching metadata from the backend, and RocksDB (\sysnameplus's backend) provides higher throughput than HDFS (\sysname's backend). For controller and server recovery, \sysname and \sysnameplus have nearly identical times, as both
procedures involve only local log reads or UDP transfers without metadata fetching.